\documentclass[english]{article}
\usepackage[T1]{fontenc}
\usepackage[left=4cm,top=4cm,right=3cm,bottom=3cm]{geometry}
\pdfoutput=1
\usepackage{float}
\usepackage{amsmath}
\usepackage{amssymb}
\usepackage{subcaption}
\usepackage[labelformat=parens,labelsep=quad,skip=3pt]{caption}
\usepackage{graphicx}
\usepackage{setspace}
\usepackage{lineno}
\usepackage{multirow}
\usepackage{hyperref}
\usepackage{soul}
\usepackage{algorithm}
\hypersetup{
    colorlinks=true,
    linkcolor=blue,
    filecolor=magenta,      
    urlcolor=cyan,
}
\makeatletter

\usepackage{beamerarticle,pgf}

\newcommand\makebeamertitle{\frame{\maketitle}}
\AtBeginDocument{
	\let\origtableofcontents=\tableofcontents
	\def\tableofcontents{\@ifnextchar[{\origtableofcontents}{\gobbletableofcontents}}
	\def\gobbletableofcontents#1{\origtableofcontents}
}

\makeatother

\usepackage{babel}
\newtheorem{remark}{Remark}
\newtheorem{assumption}[theorem]{Assumption}

\newcommand{\A}{\mathbf{A}}
\newcommand{\C}{\mathbf{C}}
\newcommand{\U}{\mathbf{U}}
\newcommand{\V}{\mathbf{V}}
\newcommand{\E}{\mathbf{E}}
\newcommand{\Vh}{\widehat{\V}}

\newcommand{\Ah}{\widehat{\A}}
\newcommand{\Ch}{\widehat{\C}}

\newcommand{\W}{\mathbf{W}}
\newcommand{\K}{\mathbf{K}}
\newcommand{\Hb}{\mathbf{H}}

\newcommand{\G}{\mathbf{G}}
\newcommand{\F}{\mathbf{F}}
\newcommand{\Y}{\mathbf{Y}}
\newcommand{\y}{\mathbf{y}}
\newcommand{\e}{\mathbf{e}}

\newcommand{\R}{\mathbb{R}}

\newcommand{\cF}{\mathcal{F}}
\newcommand{\cH}{\mathcal{F}}
\newcommand{\cN}{\mathcal{N}}
\renewcommand{\cH}{\mathcal{H}}

\newcommand{\sL}{\mathsf{L}}
\newcommand{\sN}{\mathsf{N}}

\begin{document}
\title{Error control in the numerical posterior distribution in the Bayesian UQ analysis of a semilinear evolution PDE}
\author{Maria L. Daza-Torres\footnotemark[1]\ \footnotemark[3]
\and
J. Cricelio Montesinos-L\'opez \footnotemark[1]
\and
Marcos A. Capistr\'an\footnotemark[1]
\and 
J. Andr\'es Christen\footnotemark[1]
\and 
Heikki Haario\footnotemark[2]
}

\renewcommand{\thefootnote}{\fnsymbol{footnote}}

\footnotetext[1]{Centro de Investigaci\'on en Matem\'aticas (CIMAT), Jalisco S/N, Valenciana, Guanajuato,
36023, M\'exico.
\textit{mdazatorres, jose.montesinos, marcos, jac at cimat.mx}}
\footnotetext[2]{Lappeenranta University of Technology, Department of computational and process engineering, Lappeenranta, Finland and Finnish Meteorological Institute, Helsinki, Finland \textit{heikki.haario@lut.fi}}

\footnotetext[3]{Corresponding author}

%%%%%%%%%%%%%%%%%%%%%%%

\makebeamertitle

We elaborate on results obtained in \cite{christen2018} for controlling the numerical posterior error for Bayesian UQ problems, now considering forward maps arising from the solution of a semilinear evolution partial differential equation. Results in \cite{christen2018} demand an estimate for the absolute global error (AGE) of the numeric forward map. Our contribution is a numerical method for computing the AGE for semilinear evolution PDEs and shows the potential applicability of \cite{christen2018} in this important wide range family of PDEs. Numerical examples are given to illustrate the efficiency of the proposed method, obtaining numerical posterior distributions for unknown parameters that are nearly identical to the corresponding theoretical posterior, by keeping their Bayes factor close to 1.

%\keywords{Uncertainty Quantification, Inverse problems, Bayesian inference, Evolution PDEs, Markov Chain Monte Carlo, Error estimation in PDEs}

\section{Introduction}

A wide range of applications are concerned with the  solution of an inverse problem (IP) \cite{kaipio2011bayesian, cai2011bayesian, chadan1997introduction, holder2004electrical, burggraf1964exact, snieder1999inverse, daza2017solution}: given some observations of the output, $\mathbf{y} = (y_1, \ldots, y_n)$, to determine the corresponding inputs $\theta$  such that
$$
y_i = \cF (\theta) + ~\text{error}.
$$

We refer to the evaluation of $\cF$ as solving the forward problem, and consequently, $\cF$ is called the Forward Map (FM). In general, the FM is a complex non-linear map, with input parameters  $\theta$, defined by an initial/boundary value problem for a system of ordinary differential equations (ODEs) or partial differential equations (PDEs). Then, to evaluate $\cF(\theta)$, we must solve an initial/boundary value problem for a system of (O, P)DEs.

IPs are typically ill-posed: there may be no solution, or the solution may not be unique and may depend sensitively on $y_i$ \cite{kaipio2006statistical}. A way to approach these difficulties is to formulate the IP in the Bayesian framework. Stuart \cite{stuart2010inverse} studied conditions for the well-posedness of the Bayesian formulation of IPs. In this scheme, a noise model is assumed for the observations, e.g.,
$$
y_i = \cF (\theta) + \varepsilon_i ; \quad \varepsilon_i \sim \cN (0, \sigma^2).
$$
This observational model generates a probability density given the parameter $\Phi =(\theta,\sigma)$, namely $P_{\Y |\Phi}\left(\boldsymbol{y}|\theta,\sigma\right)$, for fixed data $\mathbf{y}$, obtaining the likelihood function. Based on the available information, a prior model $P_{\Phi}(\cdot)$ is stated for $\Phi$, and a posterior distribution is obtained,
$$
P_{\Phi | \Y} \left( \theta, \sigma| \y \right)  = \frac{P_{\Y | \Phi} \left( \y | \theta, \sigma \right) P_{\Phi} \left( \theta, \sigma \right)} {P_\Y \left( \y \right)}.
$$

Explicit analytic forms are usually not available for the posterior distributions, so sampling approaches such as the Monte Carlo Markov Chain (MCMC) are required to characterize it. These methods involve repeated FM solutions used to define the likelihood function.

Usually, we do not have an analytical or computationally precise and straightforward implementation of the FM. This necessarily involves a numerical approximation, $\cF^{ \alpha ( n )}$, where $\alpha(n)$ represents a discretization used to approximate the FM,  leading to a numerical/approximate posterior distribution. Thus, the numerical solution of the FM will introduce some numerical error in the posterior distribution. At least theoretically, numerical errors in the FM can be controlled and reduced to an arbitrarily low level, through the use of finer discretizations, but what numerical error must be tolerated in the FM to obtain a correct and acceptable numerical posterior distribution?

Several approaches start by building cheap computationally approximations of the FM and using these approximations as surrogates in the sampling procedure \cite{marzouk2007stochastic, galbally2010non, lieberman2010parameter, rasmussen2003gaussian}. Although such approaches can be quite effective at reducing computation cost, there has been little analysis of posterior inference approximation. Recently, adaptive multi-fidelity techniques have been developed to control the numerical posterior error for Bayesian UQ \cite{cui2015data, peherstorfer2018survey, yan2019adaptive, yan2019adaptivePC, li2014adaptive}. In \cite{li2014adaptive} proposed an adaptive multi-fidelity polynomial chaos (PC) MCMC algorithm to find a distribution that is ``close'' to the posterior in the sense of Kullback-Leibler divergence. Similar approaches were proposed in \cite{yan2019adaptivePC, li2014adaptive} using an adaptive multi-fidelity PC based ensemble Kalman inversion technique.

Close in spirit to the works mentioned, in \cite{Capistran2016} proposes the use of Bayes Factors (BF; the odds in favor) of the numerical model vs the theoretical model. In an ODE framework, they show that the BF converge to 1, that is, both models would be equal, in the same order as the numerical solver used.

Later, this idea was generalized in \cite{christen2018} to consider also PDEs and, more importantly, the use of the \textit{expected} value of the BFs, before observing data. This results in more practical and workable guidelines in a more realistic multidimensional setting. The main result in \cite{christen2018} is a bound for the expected BF. This bound allows deciding what precision to run the solver, which could require less computational effort. Indeed, a reliable estimate of the error for the numerical method used is the central point in the calculation of this bound.

Current efforts to estimate the discretization error focus on \textit{after-the-fact} methods (i.e. \textit{a-posteriori} methods, we prefer to call then \textit{after-the-fact} to avoid the obvious confusion with the Bayesian jargon). These methods provide an error estimate only after the numerical solution has been computed. They use the computed solution to the discrete equations, possibly with additional information supplied by the equations, to estimate the error relative to the exact solution of the mathematical model \cite{roy2010review}. Most of the previous works are based on higher error bounds with asymptotic convergence when the mesh size tends to zero \cite{babuvska1978posteriori, rognes2013automated, de1983posteriori, ainsworth2011posteriori}. Unfortunately, these estimates imply ``constants of stability'' generally unknown and difficult to calculate. The resulting error estimation techniques, in practice, do not provide mathematically proven bounds that, in general, can be computed efficiently \cite{gratsch2005posteriori}.

In this paper, we derive an after-the-fact error estimate for a numerical approximation of the physical models involving a semi-linear evolution differential equation of the form:
\begin{equation}
\frac{\partial u}{\partial t} = D \frac{\partial^2 u}{\partial x^2} +F \left( u, \frac{\partial u}{\partial x} , \theta \right),\label{eq:PDE}
\end{equation}
defined on the region $t\in [0,\tau],\quad x \in [a,b]$, with left and right boundary conditions
\begin{equation}\label{eq:bound_condi}
u(a,t)  =  g(t) ~~\text{and}~~ u(b,t)  =  h(t), \quad 0\leq t \leq \tau,
\end{equation}
and initial condition
\begin{equation}\label{eq:initial_cond}
u(x, 0) = f(x), \quad a \leq x \leq b.
\end{equation} 
In Eq.~\eqref{eq:PDE}, $D$ is the diffusion coefficient, $\theta$ is a parameter (possibly a vector) of interest, and $F$ is a non-linear operator.

This physical model arises in several fields of science and engineering \cite{tzafestas2013distributed, diagana2018semilinear}. It is used to describe many complex nonlinear settings in applications such as vibration and wave propagation, fluid mechanics, plasma physics, quantum mechanics, nonlinear optics, solid-state physics, chemical kinematics, physical chemistry, population dynamics, and many other areas of mathematical modeling.

The numerical solution for Eq.~(\ref{eq:PDE}) is obtained by discretizing first in the space with the finite difference (FD) method and solving the resulting system in time with the Runge-Kutta Cash-Karp (RKCK) method \cite{cash1990variable}. This scheme is widely used to solve numerically evolution partial differential equations \cite{dlamini2017new, bastani2012highly, sari2011sixth}. However, numerical after-the-fact error estimates for these methods have not yet been derived.
 
The idea behind our construction of the error estimates for the PDE in Eq.~(\ref{eq:PDE}) is the available error estimates for the RKCK method. Our numerical method uses these error estimates, in time, for the resulting ODE system. The truncation error introduced for the approximation with finite differences is computed using the solutions in two different mesh sizes.  In modern computers, the added computational effort can be reduced to result equivalent to solving the PDE conventionally (on a single mesh) since evaluating the solution in two different meshes may be easily parallelized.
 
We will incorporate this after-the-fact error estimate in the result of \cite{christen2018}, for the solution of the Bayesian Inverse Problem (BIP) associated with the PDE given in Eq.~(\ref{eq:PDE}), to control the error in the posterior distribution. Numerical examples are given to illustrate the efficiency of the proposed method. We obtain numerical posterior distributions, for unknown parameters, that are nearly identical to the corresponding theoretical posterior, keeping their \textit{expected} Bayes factor close to 1.

The paper is organized as follows. In Section \ref{sec:SolNum}, we present a numerical method used for solving evolution partial differential equations numerically. In Section \ref{sec:ErrorAnaylsis}, we derive our after-the-fact error estimate for semi-linear evolution differential equations. The accuracy of our error estimate is evaluated for some classic examples. In Section \ref{sec:ECBUQ}, we propose an algorithm that incorporates the after-the-fact error estimate to control the error in the posterior distribution. Numerical examples are given in Section \ref{sec:NumericExamples} to illustrate the efficiency of the proposed Algorithm. Finally, a conclusion is given in Section \ref{sec:Con}.

%-------------------- Numerical Solution --------------------------------%
\section{Numerical Solution} \label{sec:SolNum}
Here, we introduce a common numerical procedure for the solution of semilinear evolution partial differential equations. This procedure has been widely used for solving evolution partial differential equations \cite{dlamini2017new, bastani2012highly, sari2011sixth}. The basic idea of the method is to replace the spatial derivation in the PDE with an algebraic approximation in order to obtain an ODE system. The resulting system is then solved with a standard ODE solver. We discretize in the space with the FD method and solving the ODE system with the RKCK method. We called this method FD-RKCK.

For simplicity, we denote $\dot{u} := \frac{\partial u} {\partial t}$, $u':= \frac{\partial u} {\partial x}$, and $F\left (u, u'\right)$ instead of $F \left( u, u', \theta \right)$.  Moreover, without losing generality, we can set $D = 1$ in Eq.~(\ref{eq:PDE}). We consider a one-dimensional uniform mesh, $\Omega_h$, on the region $[a,b]$, with nodes $x_{i}$, for $i=0,1,\ldots,N$, where
\begin{equation} \label{eq:Omega}
 \Omega_h: a = x_{0} < x_{1} < \cdots < x_{N} = b,
\end{equation}
and a constant step size $h$ between any two successive nodes (i.e., $h = x_{i} - x_{i-1}$).

To solve the PDE in Eq.~(\ref{eq:PDE}), we start by linearizing $F$ using the quasi-linearization method that was introduced in \cite{dlamini2017new} for solving nonlinear evolution partial differential equations. This method consists of separating the function $F$ into a linear ($\sL$) and a nonlinear ($\sN$) component, and rewriting  Eq.~(\ref{eq:PDE}) in the form
\begin{equation}\label{eq:PDE_LN}
\dot{u} = u''+ \sL[u,u'] + \sN[u,u'].
\end{equation} 
For example, in Section \ref{sec:ErrorAnaylsis} we use the Fisher equation where $F = ru(1-u)$, thus $\sL = ru$ and $\sN = -ru^2$. Afterwards, the nonlinear operator $\sN$ is approximated with a Taylor series, assuming that the difference $u_{i+1,\cdot} - u_{i,\cdot}$ and all its spatial derivatives are small. Hence
\begin{equation}\label{eq:Taylor}
\sN[u_{i+1, \cdot}, u_{i+1,\cdot}'] \approx \sN[u_{i, \cdot}, u'_{i,\cdot}] + \phi_{0, i} [u_{i, \cdot}, u'_{i, \cdot}] \cdot(u_{i+1, \cdot} - u_{i, \cdot}) + \phi_{1,i}[u_{i,\cdot}, u'_{i,\cdot}]\cdot(u'_{i + 1,\cdot} - u'_{i,\cdot}),
\end{equation}
where $u_{i,\cdot} := u(x_{i},t)$ is the solution of  Eq.~(\ref{eq:PDE}) evaluated in $(x_{i},t)$,  and  $$ \phi_{k,i}[u_{i, \cdot}, u_{i,\cdot}' ] := \frac{\partial \sN [u_{i,\cdot}, u'_{i, \cdot}]}{\partial u^{(k)}}, \quad k=0,1.$$
For simplicity, $u^{(i)}$ denotes the i-th derivative. Substituting Eq.~(\ref{eq:Taylor}) into Eq.~(\ref{eq:PDE_LN}), we get

\begin{equation}\label{eq:pde}
\dot{u}_{i+1,\cdot} \approx u''_{i + 1, \cdot} + \sL[u_{i + 1,\cdot}, u'_{i+1,\cdot}] + \sN[u_{i,\cdot}, u'_{i,\cdot}] +\phi_{0,i}[u_{i,\cdot}, u_{i,\cdot}' ] \cdot(u_{i+1,\cdot} - u_{i,\cdot}) +\phi_{1,i}[u_{i,\cdot}, u_{i,\cdot}' ] \cdot(u'_{i + 1,\cdot} - u'_{i,\cdot}), 
\end{equation}
for $i = 1,...,N-2$. Now, the spatial partial derivatives are approximated using the central difference formula. For simplicity, we use the simplest spatial derivative approximations here, while the analysis can be extended for other (e.g., five-point stencil) approximations as well,

\begin{equation}\label{eq:df_1}
u'_{i, \cdot}\approx \frac{u_{i+1,\cdot} - u_{i-1,\cdot}}{2h}, \quad
u_{i}'' \approx \frac{u_{i+1,\cdot}-2u_{i,\cdot} + u_{i-1,\cdot}} {h^2},
\end{equation}
for $i=1,\ldots,N-1$, and

\begin{equation}\label{eq:df_0}
u'_{0,\cdot} \approx \frac{u_{1,\cdot} - u_{0,\cdot}}{h}.
\end{equation}

Substituting Eqs.~(\ref{eq:df_1})--(\ref{eq:df_0}) in to Eq.~(\ref{eq:pde}), joint with the boundary condions (\ref{eq:bound_condi}) and initial condition (\ref{eq:initial_cond}), we get the following semi-discrete differential equation:
\begin{eqnarray} \label{eq:ODE-system}
\dot{\V}_h(t) & = &  \frac{1}{h^2}\A_{xx}\V_h(t) + \F(t,\V_h(t))\\ \label{eq:Cond_init}
\V_h(0) & = & \U(0)\\ \nonumber
\end{eqnarray} 
where $\V_h(t) = (v_{1,\cdot}, v_{2,\cdot}, \ldots,v_{N-1,\cdot})^T$ approximates 
\begin{equation}\label{eq:Umesh}
\U(t) = (u_{1,\cdot}, u_{2,\cdot}, \ldots,u_{N-1,\cdot})^T,
\end{equation}
$\U$ is the exact solution of the PDE \eqref{eq:PDE} on the mesh $\Omega_{h}$, and $\F$ is the approximate operator $F$ in matrix form, see \ref{sec:ap_ns} for details.

\begin{remark} The semi-discrete differential equation (\ref{eq:ODE-system}) have a truncation error $O (h^p)$:
\begin{itemize}
\item	[(i)] If $F$ does not have a nonlinear component, the quasi-linear approximation \eqref{eq:Taylor} is not necessary. Thus, the truncation error for the central difference formula is not affected ($p=2$).

\item[(ii)] If $\sN$ is non-linear in $u'$, the quasi-linear approximation \eqref{eq:Taylor} introduces a truncation error of first-order, which is propagated when $u'$ is approximated using the central difference formula. Thus, the order of the truncation error for \eqref{eq:ODE-system} is less than 2 ($p < 2$).

\item[(iii)] If $\sN$ is non-linear in $u$ and linear in $u'$, the truncation error introduced for the quasi-linear approximation \eqref{eq:Taylor} is not propagated as in case (ii). Then, the truncation error in \eqref{eq:ODE-system} is slightly affected ($p\approx 2$).
\end{itemize}
\end{remark}

In order to solve the resulting ODE system~(\ref{eq:ODE-system})-(\ref{eq:Cond_init}), with $N-2$ equations, we use the RKCK method. This method uses six function evaluations to calculate fourth and fifth-order accurate solutions. The difference between these solutions is then taken to be the error (fourth-order) of the solution; see \cite{burden2000numerical} for details. The available error estimate is the reason to solve the resulting ODE's system with this Runge-Kutta (RK) method, and it will be used in turn, in Section \ref{sec:ErrorAnaylsis}, for computing the after-the-fact error of the numerical solution of Eq.~(\ref{eq:PDE}).

Setting $\G (t, \V_h(t)) = \frac{1}{h^2} \A_{xx} \V_h(t) + \F (t ,\V_h(t))$, a RK scheme applied to the ODE system~\eqref{eq:ODE-system}, at a uniform time grid
\begin{equation}\label{eq:grid_t}
0 = t_0 < t_1 < \cdots < t_n < \cdots < t_{M-1} < t_M = \tau; \quad t_{n+1} = t_{n} + k,
\end{equation}
is given by 
\begin{align} \label{eq:K1}
\K_{1,n} & = \G \left(t_n, \W_{\cdot,n}\right),\\  \label{eq:Kn}
\K_{l,n} & = \G \left(t_n + c_l k,\W_{\cdot,n} + k\sum_{j=1}^{l-1} a_{lj} \K_{j,n} \right),\quad l = 2, 3, \ldots, 6,\\  
\W_{\cdot, n+1} & = \W_{\cdot,n} + k \sum_{l=1}^6 b_l \K_{l,n}, \quad n=1,\dots, M-1, \nonumber
\end{align}
where $\W_{\cdot, n+1}$ is the approximation for $\V_h(t_{n+1})$, $(a_{lj})$ are the Runge-Kutta coefficients, $\mathbf{b} = (b_1, b_2,\ldots,b_6)$ are the quadrature nodes, and $\mathbf{c}= (c_1, c_2, \ldots, c_6)$ are the quadrature weights of the RK scheme. $k = \Delta t > 0$ is the step size in time and define a uniform grid.

In order to have stable solutions in explicit schemes, the step size in time is related to the discretization through the Courant-Friedrichs-Lewy (CFL) condition \cite{courant1928partial}, which restricts the step size in time based on the eigenspectrum of the discretized spatial operator. The CFL condition for the FD-RKCK scheme considering only the pure diffusion is 
$$
\frac{\Delta t }{\Delta x^{2}} \leq \frac{1}{4} b_{\text{max}},
$$
where $b_{max} = \max_{i} b_{i}$, and $b_1, b_2, \ldots, b_6$ are the quadrature nodes for the RK method used, see \ref{sec:ap_sc} for details.

%----------------- After-the-fact error estimates --------------------------------%
\section{After-the-fact error estimates}\label{sec:ErrorAnaylsis}
In this section, we propose a numerical procedure to obtain an after-the-fact error estimate of the AGE, for the numerical solution of Eq.~(\ref{eq:PDE}). For them, we use the error estimation in the time-stepping given for the RKCK method and estimate the leading term of the truncation error in space stepping. This scheme can be extended for differential equations of non-linear evolution, but some additional considerations about the stability of the solution must be taken into account.

In Section \ref{sec:SolNum}, we obtained the semi-discrete differential equation \eqref{eq:ODE-system}--\eqref{eq:Cond_init}, with a unique solution vector, $\V_h(t)$, being a grid function on $\Omega_h$. This initial value problem solved with the RKCK method yields approximations $\W_{.,n}$ to $\V_h(t_n)$. The global error at the spatial mesh points at knot $t_n$ is defined by 
\begin{equation}\label{eq:error_g}
\E_h(t_n):= \W_{.,n} - \U(t_n),
\end{equation}
where $U$ is the exact solution of the PDE \eqref{eq:PDE} on the mesh $\Omega_{h}$ defined in \eqref{eq:Umesh}. The vector $E_h$ may also be written as a combination of the ODE global error, this is defined as the error made by the solver, i.e.,

\begin{equation}\label{eq:error_t}
e_{h}(t_n)=\W_{.,n} - \V_{h}(t_{n}), 
\end{equation}
and the spatial discretization error defined by

\begin{equation}\label{eq:error_h}
\eta_h(t_n) =\V_h(t_n)-\U(t_n). 
\end{equation}

The function $\eta(t)$ represents the accumulation of the spatial truncation error (TE) when we solve \eqref{eq:ODE-system}-\eqref{eq:Cond_init},

\begin{equation}\label{eq:TE}
TE_{h}(t) = \G(t,\U)-\dot{\U}(t).
\end{equation}

From Eqs.~(\ref{eq:error_t})--(\ref{eq:error_h}), the global error $\E_h(t_n)$ may be written as the sum of the global time and spatial error, i.e.,

\begin{equation}\label{eq:error_global}
\E_h(t_n) = e_h(t_n) + \eta_h(t_n).
\end{equation}

We assume that $u(t,x)$ is $p$-times differentiable with respect to $x$ and fourth-times continuously differentiable with respect to $t$. Then, it holds for the global space and time error that $||\eta_h|| = O(h^p)$ and $||e_h(t_n)|| = O(k^4)$, $n = 1, \ldots, M,$ respectively.

The ODE global error \eqref{eq:error_t} is calculated using the error estimation of RKCK \cite{dlamini2017new}. The spatial discretization error implementation based on \eqref{eq:eta_ode}-\eqref{eq:eta_cond} requires an estimation for the truncation error. The Richardson extrapolation \cite {roy2010review} provides a suitable estimate of the truncation error. The idea is to calculate the solution using a one-step size $h$  and then compute them again with half the space step ($h / 2$). The result obtained using two steps size is more accurate than using the single-step size h. Their difference can be used as an estimate of the truncation error, which is proportional to the power of $h$.

\subsection{Spatial discretization error}
We can obtained an equation for the evolution of $\eta(t)$ by adding terms to both sides of \eqref{eq:ODE-system}:
$$
\dot{\V}_h(t)-\dot{\U}(t_{n}) = \G(t,\V_h)-\G(t,\U)+ \G(t,\U)-\dot{\U}(t_n).
$$
From the initial condition \eqref{eq:Cond_init} and using the definition $\eta(t)$ in the above equation, the accumulation of the spatial discretization error is the solution to the initial value problem: 
\begin{eqnarray}\label{eq:eta_ode}
\dot{\eta}(t)	&=&	\G(t,\V_h) - \G(t,\U) + TE_h(t),\quad t\in(0,\tau]\\
\eta(0)	&=&	0.\label{eq:eta_cond}
\end{eqnarray}
Assuming $\G$ to be twice continuously differentiable, we use the approximation:
\begin{equation}\label{eq:dG}
\frac{\partial \G}{\partial \V_h}\approx\frac{\G(t,V_{h}) - \G(t, \U)}{\V_h - \U}.
\end{equation}
Finally, we rewrite \eqref{eq:eta_ode}-\eqref{eq:eta_cond} to get

\begin{eqnarray}\label{eq:eta_ode1}
\dot{\eta}(t)	&=&	\frac{\partial G}{\partial V_{h}}\eta(t) + TE_h(t),\quad t\in(0,\tau]\\
\eta(0)	&=&	0. \label{eq:eta_cond1}
\end{eqnarray}

The integration of \eqref{eq:eta_ode1}--\eqref{eq:eta_cond1} is performed  using $M$ steps of size $k$ of the RKCK method, as in the solution of the semi-discrete differential equation \eqref{eq:ODE-system}--\eqref{eq:Cond_init}. In each RKCK step, $\frac{\partial \G}{\partial \V_h}$ is approximated using the approximations $\W_{.,n}$ and $\W_{.,n+1}$ to $\V_h$ at time $t_{n+1}$, i.e.,

\begin{equation}
\frac{\partial \G}{\partial \V_h} \approx \frac{\G(t_{n+1},\W_{.,n+1}) - \G(t_{n+1}, \W_{.,n})}{\W_{.,n+1}-\W_{.,n}}.
\end{equation}

\subsection{Spatial and time error}
The ODE global error \eqref{eq:error_t} is computed by the error estimation given by the RKCK method. This scheme uses an RK method with a fifth-order local truncation error to estimate the local error in an RK method of fourth-order. Both with the same number of stages $s=6$, Runge Kutta matrix $\mathbf{A}$, and weights $\mathbf{c}$, while their nodes $\hat{\mathbf{b}}$ and $\mathbf{b}$, respectively, are different; see \cite{dlamini2017new} for details. 

Let $\mathbf{W}_{\cdot , n+1}$ the $n+1$ approximation of $\V_h(t_{n+1})$ of fourth-order, and let $\mathbf{Y}_{\cdot , n+1}$  be obtained by the fifth-order method starting at $\W_{\cdot, n}$, namely

\begin{equation}\label{eq:solU}
\W_{\cdot , n+1} = \W_{\cdot, n} + k \sum_{i=1}^s b_i \K_{i,n} \quad \text{and}
\quad \Y_{\cdot, n+1} = \W_{\cdot, n} + k\sum_{i=1}^s \hat{b}_{i} \K_{i,n},
\end{equation}

The local truncation error $\hat{\mathbf{\tau}}_{\cdot, n+1}$ at node $t_{n+1}$ of the RK method is defined as the error made in step $n + 1$ of the solver if starting at the exact value $\W_{\cdot, n}$. The estimation of $\hat{\mathbf{\tau}}_{\cdot, n+1}$ for the RKCK method is given by 
$$
\hat{\mathbf{\tau}}_{\cdot, n+1} = \mathbf{Y}_{\cdot,n+1} - \mathbf{W}_{\cdot,n+1} = k  \sum_{i=1}^{s} (\hat{b}_{i} - b_{i})\mathbf{K}_{i,n},
$$
and the global error at knot $t_{n+1}$ is
\begin{equation}\label{eq:ErrorEst}
\hat{e}_{., n+1} = \sum_{j=1}^{n+1} \tilde{\tau}_{\cdot,j}. 
\end{equation}
In each RK iteration, we solve the equation for the spatial discretization error \eqref{eq:eta_ode1}--\eqref{eq:eta_cond1},

\begin{equation}\label{eq:ODE-spatial-error-dis}
\hat{\eta}_{n+1} = \hat{\eta}_n + k\sum_{i=1}^s b_i\hat{K}_{i,n},
\end{equation}
where

\begin{eqnarray*}
\hat{K}_{1,n} & = & H(t_n,\hat{\eta}_{n})\\
\hat{K}_{l,n} & = & H\left(t_n+c_{l}k,\eta_{n}+k\sum_{j=1}^{l-1}a_{lj}\hat{K}_{j,n}\right),\quad l=2,3,\ldots,6.
\end{eqnarray*}
$H$ is the right side of \eqref{eq:eta_ode1}. From \eqref{eq:ErrorEst}--\eqref{eq:ODE-spatial-error-dis}, an estimation for the global error $\E_h$ \eqref{eq:error_global} at knot $t_n$ is given by,

\begin{equation}\label{eq:Eglobal}
\hat{\E}_{.,n+1}\approx \hat{e}_{.,n+1} + \hat{\eta}_{n+1}.
\end{equation}

Note that to solve fully \eqref{eq:eta_ode1}--\eqref{eq:eta_cond1}, we need an estimate for the truncation error. This estimation is done in parallel to be used in \eqref{eq:ODE-spatial-error-dis}. Below we give details for computing the truncation error.

\subsection{Spatial truncation error}  

The truncation error is the difference between the discretized equations and the original partial differential equations. It contains the errors due to the discretization of the PDE and the errors due to the grid. For the finite difference scheme used to approximate the spatial operator, we have that the truncation error at time $t$ has rate order $O(h^{p}),$
$$
TE_{h}(t)\approx O(h^{p}).
$$

An efficient strategy to estimate the spatial truncation error by Richardson extrapolation is proposed in \cite{roy2010review}. We will adopt this approach to our setting. The actual mesh used to compute the numerical solution to the PDE is used as the fine mesh in the Richardson extrapolation process. Suppose we are given a second semi-discretization of the PDE system \eqref{eq:PDE}, now with doubled local mesh sizes defined as follows,

$$\Omega_{2h}:=a=z_{0}<z_{1}<z_{2}<\ldots<z_{N/2}=b,\quad z_{i}=x_{2i},\quad i=0,\ldots,N.$$

This mesh is called the coarse mesh.
We assume that the solution $V_{2h}(t)$ to the discretized PDE, on the coarse mesh $2h$, exists and is unique. The Richardson extrapolation gives an estimation of the truncation error for the fine mesh at time $t$,

\begin{equation} \label{eq:sp_error}
\widehat{TE}(t) \approx  \frac{ \V_{2h} - R_{2h}(\V_h)} { 2^p - 1},
\end{equation}	
where $R_{2h}$ is the usual restriction operator defined by

$$
R_{2h}(V_h) = \left(v_{1,.}, v_{2,.}, \ldots,v_{(N-1)/2,.} \right)^T, \quad v_{i,.}=v(z_i,t), \quad z_i \in \Omega_{2h}.
$$ 

\begin{remark}
In the compute of \eqref{eq:sp_error}, we have found in computational experiment that this term could be approximate (on $\Omega_{h} \times (0,\tau]$) at time $t$ as
$$
\widehat{TE}_{h}(t) = \frac{h^{p}}{\tau}.
$$
This estimation is valid for our settings, but is not applicable if another scheme for solving \eqref{eq:PDE} is used.
\end{remark}
\begin{remark} The after-the-fact error estimate of the AGE for the FD-RKCK solution of \eqref{eq:PDE}, on $\Omega_h \times (0, \tau]$, is given by 
$$
\hat{K} = || \hat{\mathbf{E}}_{., M} ||_{\infty},
$$
where $\Omega_h$ is defined in \eqref{eq:Omega}, the discretization grid for $(0, \tau]$ is defined in \eqref{eq:grid_t}, and $\hat{\mathbf{E}}_{., M}$ is an estimate for the global error \eqref{eq:error_global}.
\end{remark}

In Algorithm \ref{alg:num_est}, we describe the steps necessary to compute the numerical solution of Eq.~(\ref{eq:PDE}), with the after-the-fact error estimation. We call this algorithm DF-RKCK.

\begin{algorithm}[ht] 
	\caption{DF-RKCK} %Numerical solution and error computing
	\label{alg:num_est}      
\textit{Step 1: }{\bf Initialization:}\\\vspace{-0.5cm}
\begin{itemize}
\item  Spatial step size $h$. The step size in time is given for keeping the stability condition $k=\alpha h^{p}$
\item Initial conditions $W^{h}_0$ and $W^{2h}_0$; initial time $t_0$; parameter $\theta$; $\hat{\e}_{.,0} = \mathbf{0}$ and $\widehat{TE}=0$
\item The RK matrix $\mathbf{A} = \left(a_{ij} \right)$, the nodes $\mathbf{b}$ and $\hat{\mathbf{b}}$, and the weights $\mathbf{c}$ 
\end{itemize}

\textit{Step 2.} Discretizing~(\ref{eq:PDE}) with the FD method for $h$ and $2h$, as is described in Section  \ref{sec:SolNum}.\\

\textit{Step 3.} Solve (\ref{eq:ODE-system}) with the Cash-Karp method for the step size $h$ and $2h$:\\

\hspace{8mm} For $n = 1, 2, \ldots, M$:\\ \vspace{-.6cm}
\begin{align*}
		\K^{h}_{1,n}       & = \G \left(t_{n}, \W_{\cdot, n} \right); \qquad \hat{\K}_{1,n} = \Hb \left(t_{n}, \hat{\eta}_{n}\right)\\
		\K^{2h}_{1,n}       & = \G \left(t_{n}, \W_{\cdot, n} \right)
\end{align*}

\vspace{-0.1cm} \hspace{14mm}
\textit{Step 4.} For $i=2, 3, \ldots, 6$: \\ \vspace{-.6cm}
	\begin{align*}
       	\K^{h}_{i,n}       & = \G \left(t_{n} + c_{i}k, \W_{\cdot,n} + k \sum_{j=1}^{i-1} a_{ij}\K^{h}_{j,n} \right)\\
       	\hat{\K}_{i,n} & = \Hb \left(t_n + c_{i}k, \hat{\eta}_{n} + k\sum_{j=1}^{i-1} a_{lj} \hat{\K}_{j,n}\right)\\
       	\K^{2h}_{i,n}       & = \G \left(t_{n} + c_{i}k, \W_{\cdot,n} + k\sum_{j=1}^{i-1} a_{ij}\K^{2h}_{j,n} \right)
\end{align*}
\vspace{-0.5cm}

\hspace{14mm} \textit{Step 5.} Compute \vspace{-0.5cm}
\begin{align*}
\W^{h}_{\cdot,n+1}  & = \W^{h}_{\cdot,n}  + k \sum_{i=1}^{6} b_{i}\K^{h}_{i} ; \quad \hat{\eta}_{n+1} = \hat{\eta}_n + k \sum_{i=1}^6 b_i \hat{\K}_{i,n}\\
\W^{2h}_{\cdot,n+1} & = \W^{2h}_{\cdot,n} + k \sum_{i=1}^{6} b_{i}\K^{2h}_{i}; \quad \widehat{TE}  = \frac{\left\Vert R_{2h}(\W^{h}_{\cdot,n+1}) - \W^{2h}_{\cdot,n+1} \right\Vert _{\infty}}{h^{p} \left(2^{p}-1 \right)}\\
\hat{\e}_{.,n+1} & = \hat{\e}_{.,n} + k \sum_{i=1}^6 (b_{i} - \hat{b}_{i}) \K^{h}_{i}\\
\widehat{\E}_{.,n+1} & = \hat{\e}_{.,n+1} + \hat{\eta}_{n+1}
\end{align*}

\textit{Step 6.} Compute the maximum absolute global error in the solution approximated $\mathbf{\W}^{h},$ 
$$
\widehat{K} =  \left\Vert \widehat{\E}  \right\Vert _{\infty}
$$

\textit{Step 7:} Output: $\mathbf{W}^{h}$, $\widehat{K}$
\end{algorithm}

\begin{figure}[ht]
\centering
\includegraphics[scale=0.3]{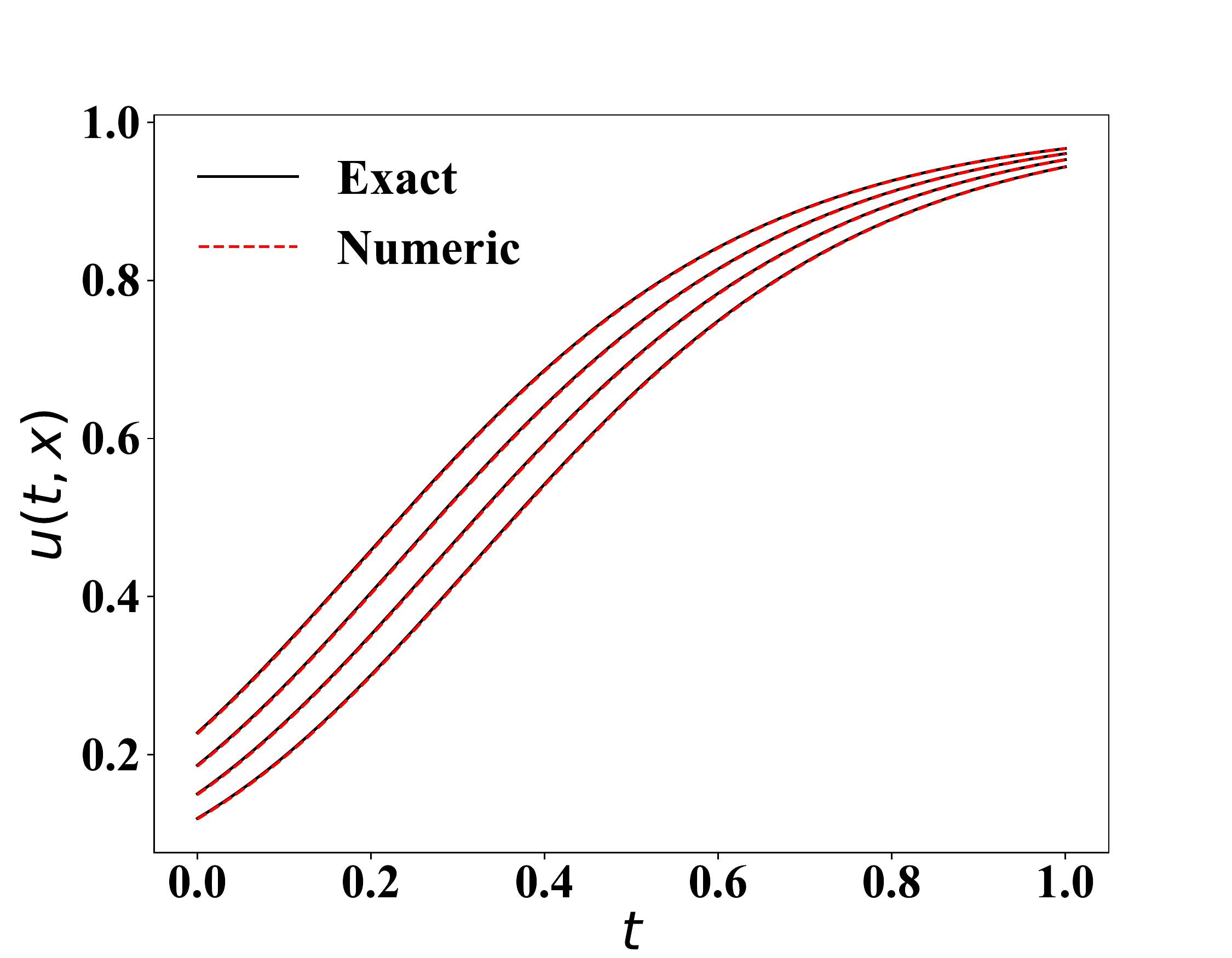}\includegraphics[scale=0.3]{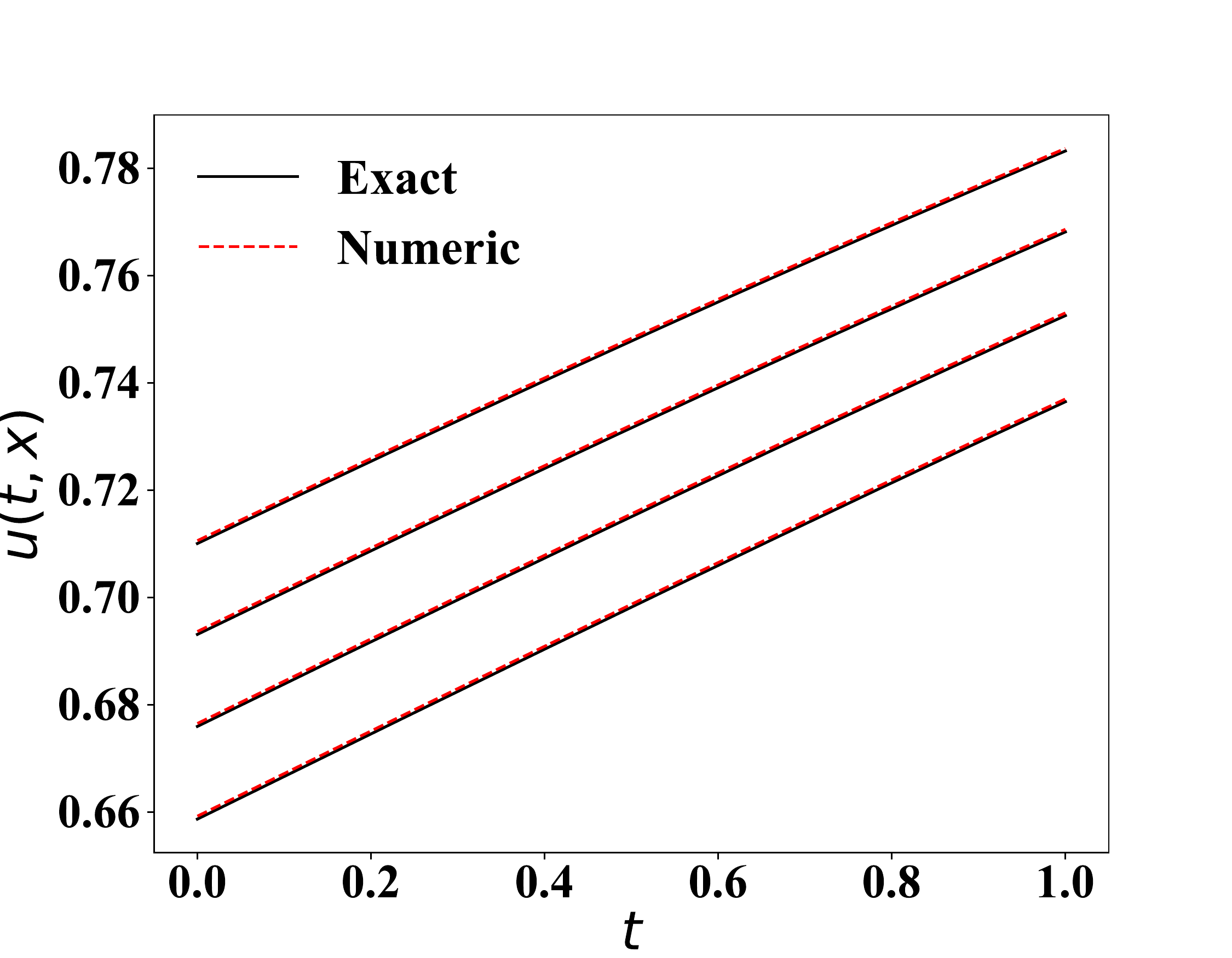}\\
(a)\hspace*{200pt}(b)\\
\centering
\includegraphics[scale=0.3]{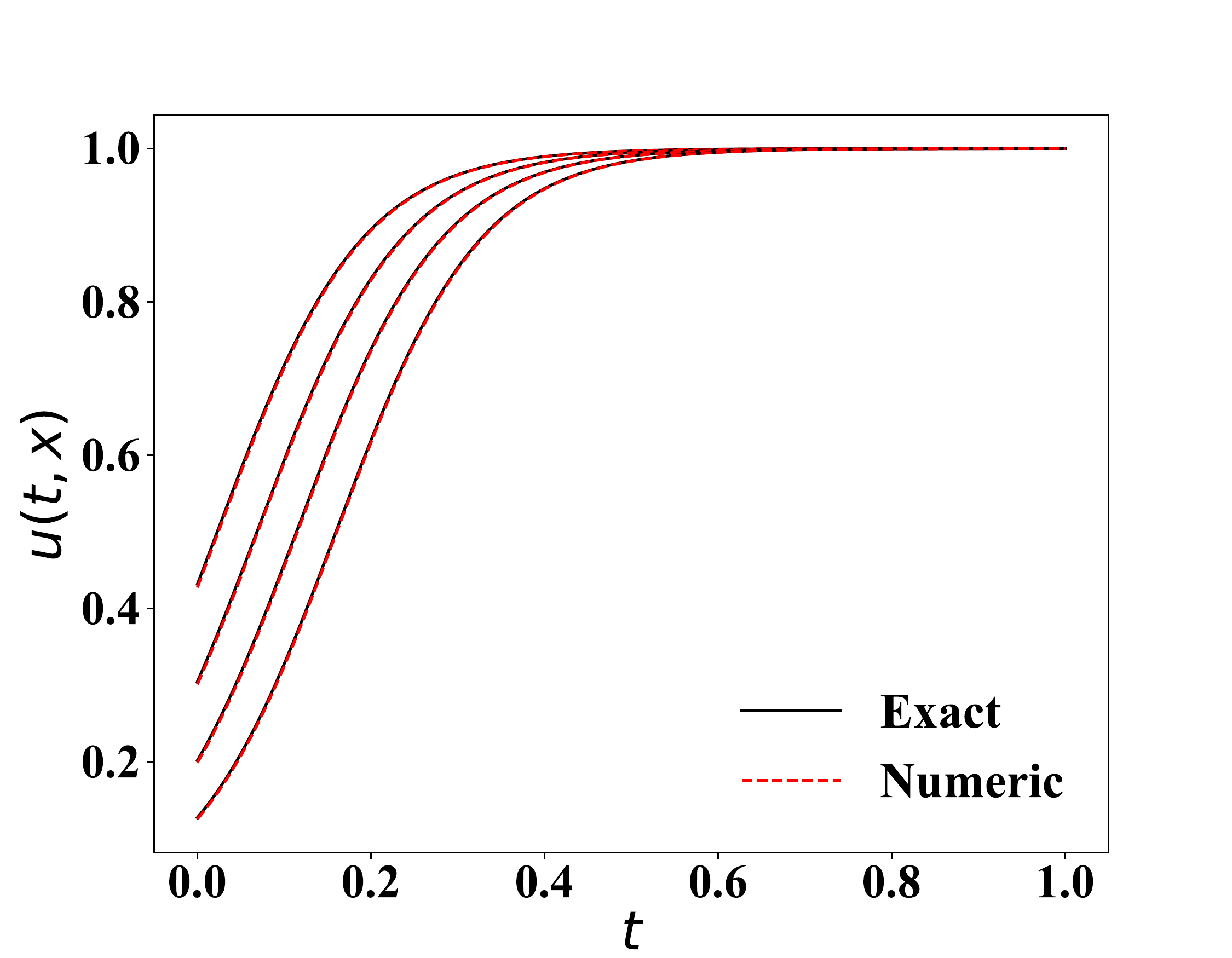}\\
(c)
\caption{The analytical and approximate solution in $x = 0.1, 0.3, 0.5, 0.7$, with step sizes in space $h=0.0125$ and in time $k=0.0001$,  for (a) the Fisher's equation, (b) the Fitzhugh-Nagumo equation, and (c) the Burgers-Fisher equation.
 \label{fig:sol_fisher}}
\end{figure}

To test our algorithm, we consider three classical semi-linear PDEs, of the form (\ref{eq:PDE}): Example \ref{ex:fisher} (Fisher equation), Example \ref{ex:Fitz-Nag} (Fitzhugh-Nagumo equation), and Example \ref{ex:Burgers} (Burgers-Fisher equation).  The three examples used also have analytic solutions, allowing us to compute the actual numerical error and compare it with our estimates.
In Figure \ref{fig:sol_fisher}, a graphical comparison is shown between our numerical implementation approximations and the exact solution for the three examples.  Table \ref{tab:order} shows the convergences order of the solution obtained with the  DF-RKCK Algorithm. It can be seen that the method achieves full convergence for the error (order 2) for Examples \ref{ex:fisher} and \ref{ex:Fitz-Nag}, but the order of convergence for Example \ref{ex:Burgers} is $1$, and this is due to the non-linearity of $F$ in $u'$, as was mentioned before.

\begin{remark}
To compute the numerical convergence rate, we use 
$$
p= \log_{2}\left(\frac{||u_{4h} - u_{2h}||_{\infty}}{||u_{2h} - u_{h}||_{\infty}}\right).
$$
\end{remark}

\begin{table}[ht]
\centering
\caption{Convergence Order\label{tab:order}}
\centering\begin{tabular}{|c|ccc|}
\hline
\textbf{h}   & \textbf{Example 1} & \textbf{Example 2} & \textbf{Example 3} \\
\hline
0.0125 &  1.999092 &  1.992578 &  1.075049 \\
0.0083 &  1.999617 &  1.995410 &  1.052625 \\
0.00625 &  1.999789 &  1.996692 &  1.040450 \\
0.005 &  1.999866 &  1.997418 &  1.032826 \\
\hline
\end{tabular}
\end{table}

\begin{figure}[ht]
\centering
\includegraphics[scale=0.3]{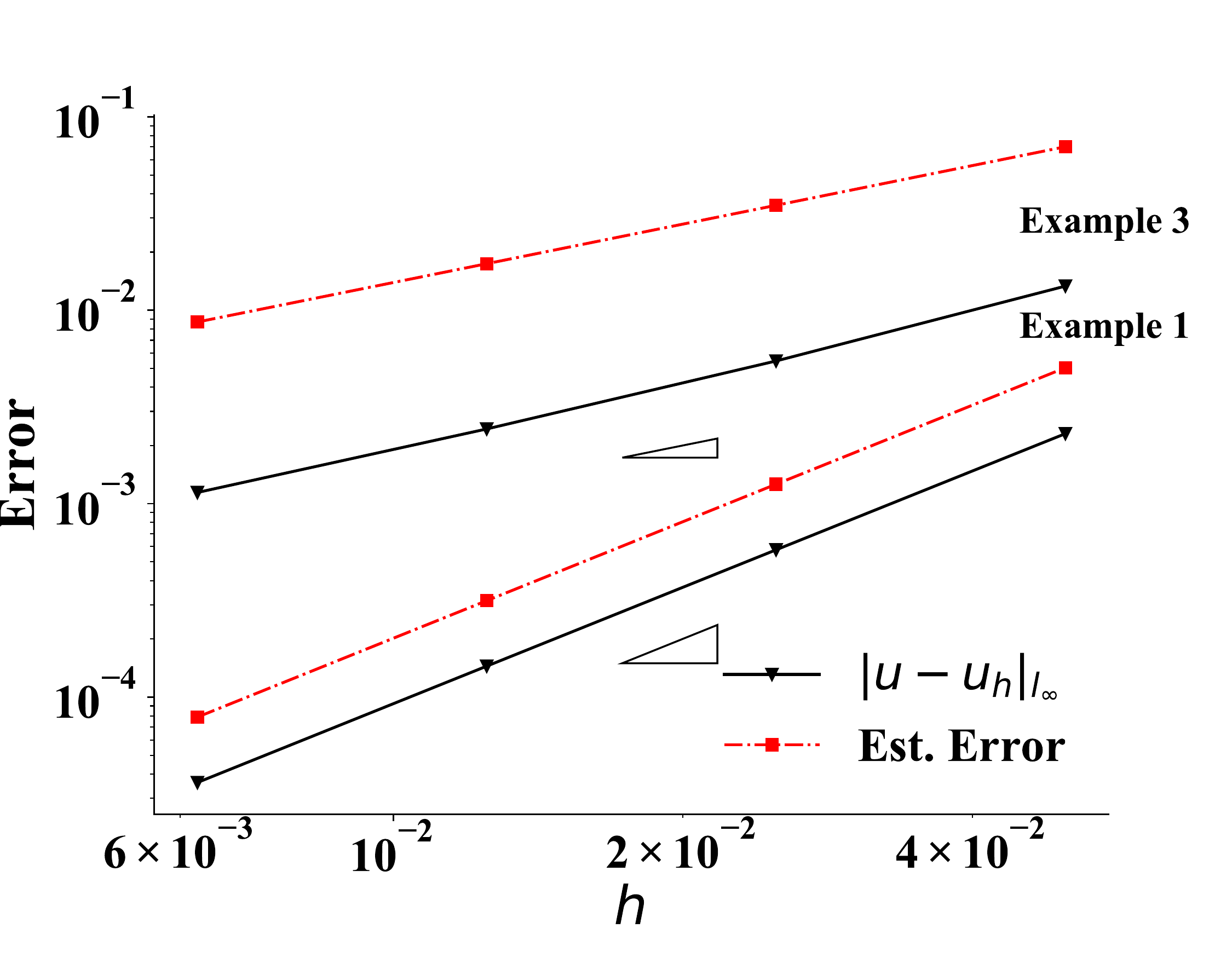}\includegraphics[scale=0.3]{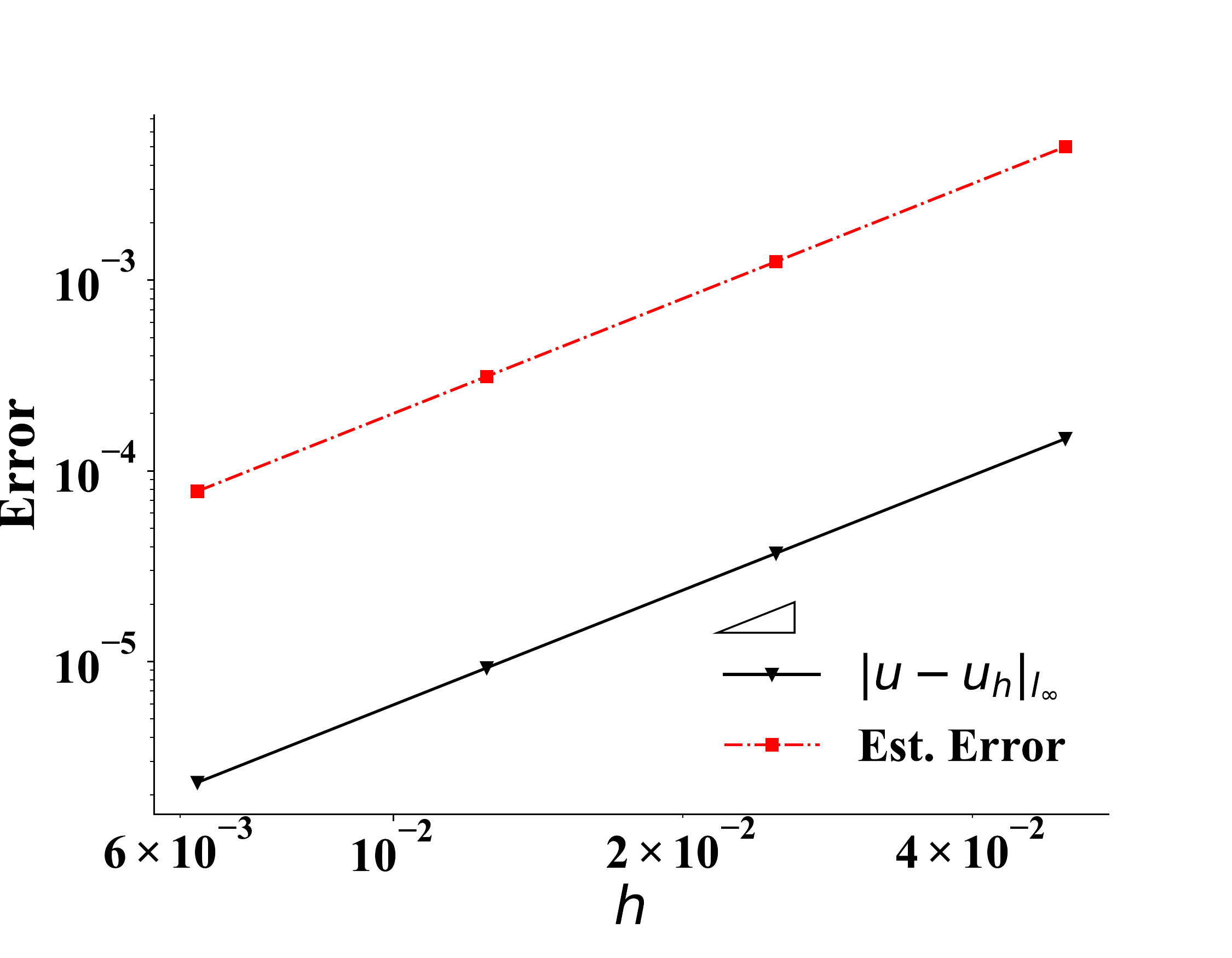}\\
(a)\hspace*{200pt}(b)
\caption{The maximum error between the exact solution and our DF-RKCK method
against the error estimation: (a) the Fisher and the Burgers-Fisher equation, (b) the Fitzhugh-Nagumo equation. Different step sizes ($h$) in space are taken and for time we let $k = \alpha h^{2}$, with $\alpha=3/4$.\label{fig:error_est} }
\end{figure}

In Figure \ref{fig:error_est}, we show the maximum error between the exact solution and the numerical solution for the three examples considered, comparing it to our error estimates. We can see that the estimation proposed for the absolute global error is an upper bound for the exact error.  The numerical implementation has been performed in Python, using the scipy, numpy, and matplotlib packages. For the sake of reproducibility, all code is available in a Github repository \cite{codes}.

\begin{example}[Fisher equation]\label{ex:fisher}
Fisher's equation belongs to the class of reaction-diffusion equation  and is encountered in chemical kinetics and population dynamics applications. The equation is given by
\begin{equation}\label{eq:Exa1}
\frac{\partial u}{\partial t} = \frac{\partial^2 u}{\partial^2 x} + ru \left( 1-u \right),
\end{equation}
with boundary and initial conditions
\begin{eqnarray}
u(0,t) & = & \frac{1}{(1+e^{-5t})^{2}},\quad 0\leq t \leq\tau,\nonumber\\
u(1,t) & = & \frac{1}{(1+e^{1-5t})^{2}},\quad 0\leq t \leq\tau,\nonumber\\
u(x,0) & = & \frac{1}{(1+e^{x})^{2}},\quad 0\leq x \leq1 .\nonumber
\end{eqnarray}
This PDE has the following analytic close form solution
\begin{equation*}
u \left( x, t \right) = \frac{1}{\left[ 1 + \exp\left( \sqrt{\frac{r}{6}} x - \frac{5r}{6} t \right)\right]^{2}}, \quad x\in\left[0,1\right],\quad \text{and} \quad t \in \left[0,\tau\right],
\end{equation*}
where $r$ is a parameter. The non-linear operator is $F(u,u',r) = ru(1-u)$; hence the appropriate linear component is $\sL = ru$, and the non-linear component is $\sN = -ru^2$; we see that the operator $F$ does not depend on $u'$ so the method achieves order 2, as can be seen in Table \ref{tab:order}. For the examples in Figs.~\eqref{fig:sol_fisher}--\eqref{fig:error_est} and Table \ref{tab:order}, we use $r = 4$, and this parameter will be tried to identify using synthetic data in section \ref{sec:NumericExamples}.
\end{example}

\begin{example}[Fitzhugh-Nagumo equation]\label{ex:Fitz-Nag}

The Fitzhugh-Nagumo equation is given by
\begin{equation}\label{eq:Exa2}
\frac{\partial u}{\partial t} = \frac{\partial^2 u}{\partial^2 x} + u \left( 1 - u \right) \left( u - a \right), \quad 0<a<1,
\end{equation}
with boundary and initial conditions
\begin{eqnarray}
u(0,t) & = & \frac{1}{2}\left( 1 + a \right) + \frac{1}{2}\left( 1 - a \right) \tanh\left(\frac{\left(1 - a^{2} \right)}{4} t \right),\quad 0\leq t \leq\tau,\nonumber\\
u(1,t) & = & \frac{1}{2}\left( 1 + a \right) + \frac{1}{2}\left( 1 - a \right) \tanh\left( \sqrt{2} \left( 1 - a \right) \frac{1}{4} + \frac{\left(1 - a^{2} \right)}{4} t \right), \quad 0\leq t \leq\tau,\nonumber\\
u(x,0) & = & \frac{1}{2} \left( 1 + a \right) + \frac{1}{2}\left( 1 - a \right) \tanh\left( \sqrt{2}\left(1-a\right)\frac{x}{4}\right),\quad 0\leq x\leq 1 .\nonumber
\end{eqnarray}
The analytic solution for this PDE is given by
\begin{equation*}
u \left( x, t \right) = \frac{1}{2}\left( 1 + a \right) + \frac{1}{2}\left( 1 - a \right) \tanh\left( \sqrt{2} \left( 1 - a \right) \frac{x}{4} + \frac{\left(1 - a^{2} \right)}{4} t \right),\quad x\in\left[0,1\right], \quad \text{and} \quad t \in \left[0,\tau\right],
\end{equation*}
where $a$ is a parameter. The non-linear operator is $F(u,u',a) =  u \left( 1 - u \right) \left( u - a \right)$; hence the appropriate linear component is $\sL = -au$, and the non-linear component is $\sN = u^{2}\left(1-u+a\right)$; we see that the operator $F$ does not depend on $u'$, so the method achieves order 2, as can be seen in Table \ref{tab:order}. For the examples in Figs.~\eqref{fig:sol_fisher}--\eqref{fig:error_est} and Table \ref{tab:order}, we use $a = 0.3$ and this parameter will be tried to identify using synthetic data in section \ref{sec:NumericExamples}.
\end{example}

\begin{example}[Burgers-Fisher equation]\label{ex:Burgers}

The Burgers-Fisher equation is given by
\begin{equation}\label{eq:Exa3}
\frac{\partial u}{\partial t} = \frac{\partial^2 u}{\partial^2 x} - ruu' + su \left( 1 - u \right),
\end{equation}
with the initial condition
$$
u \left(x , 0 \right) = \frac{1}{2} + \frac{1}{2}\tanh\left( -\frac{r}{4} x \right), \quad 0\leq x \leq 1,
$$
and with boundary and initial conditions
\begin{eqnarray}
u(0,t) & = & \frac{1}{2} + \frac{1}{2}\tanh\left( \left( \frac{r^2}{8} + \frac{s}{2} \right) t\right),\quad 0\leq t \leq\tau,\nonumber\\
u(1,t) & = & \frac{1}{2} + \frac{1}{2}\tanh\left( -\frac{r}{4}\left[ 1 - \left( \frac{r}{2} + \frac{2s}{r} \right) t\right]\right), \quad 0\leq t \leq\tau,\nonumber\\
u(x,0) & = & \frac{1}{2} + \frac{1}{2}\tanh\left( -\frac{r}{4} x \right), \quad 0\leq x \leq 1.\nonumber
\end{eqnarray}
This problem also has an analytic solution given by
\begin{equation*}
u \left( x, t \right) = \frac{1}{2} + \frac{1}{2}\tanh\left( -\frac{r}{4}\left[ x - \left( \frac{r}{2} + \frac{2s}{r} \right) t\right]\right),\quad x\in\left[0,1\right], \quad \text{and} \quad t \in \left[0,\tau\right],
\end{equation*}
where $r$ and $s$ are parameters. The non-linear operator is $F(u,u',a) =  - ruu' + su \left( 1 - u \right)$; hence the appropriate linear component is $\sL = su$, and the non-linear component is $\sN = -ruu'- su^{2}$. Of note is that the order of convergence for the error is $1$ because $F$ is nonlinear in $u'$. For the examples in Figs.~\eqref{fig:sol_fisher}--\eqref{fig:error_est} and Table \ref{tab:order}, we use $r = 4.5$ and $s=5.5$, and these parameters will be tried to identify using synthetic data in the next section.
\end{example}

%########## Error Control in Bayesian UQ ##################
\section{Error Control in Bayesian UQ} \label{sec:ECBUQ}
In this section, we discuss how to incorporate the after-the-fact error estimate, proposed in Section~\ref{sec:ErrorAnaylsis}, in the results of \cite{christen2018}, to control the error in the posterior distribution.

We follow the general setting of \cite{christen2018} for the statistical IP. Let $\Theta$ and $V$ be separable Banach spaces. Let $\cF: \Theta \rightarrow V$ be the FM (typically $\cF(\theta)$, for all $\theta \in \Theta$,  is the solution of a system of PDE's) and $\cH: V \rightarrow A \subseteq \R^{m}$ the observation operator (e.g., $\cH \left( \cF(\theta) \right)$ is one particular state variable, for which we have observations). The composition $\cH \circ \cF$ defines a mapping from the parameter space $\Theta$ to the data sample space in $\R^m$. Also, assume that $f( \y | \theta)$ is a density for data $\y$:
$$
f( \y | \theta) := f_o( \y | \cH( \cF( \theta ) ) ); \quad \theta \in \Theta,
$$
where $f_o(y | \eta (\theta ))$ is a density function that interacts with $\theta$ only through $\eta (\theta ) \in \R^m $.

Let $\cF^{\alpha(n)}$ be a discretized version of the FM $\cF$, for some discretization $\alpha$ that depends on an integer refinement $n$, e.g., a spatial step size in FD discretization. And, let $f^{n}( \y | \theta) :=  f_o( \y | \cH(\cF^{\alpha(n)}(\theta)) )$ be the resulting discretized numerical likelihood.

To find reasonable guidelines, to choose a discretization level, in \cite{Capistran2016} compare the numeric posterior with the theoretical posterior using Bayesian model selection, namely Bayes Factors (BF). Assuming an equal prior probability $\pi$ for both models, the BF is the ratio of the normalization constants $\frac{Z^n( \y )}{Z( \y )}$, where
$$
Z( \y ) = \int f(\y | \theta) \pi(\theta) d\theta,
$$
and $Z^n( \y )$ is the corresponding numeric normalization constant.

Later, in \cite{christen2018} try to control the BF between the discretized model and the theoretical model, through the use of the Absolute BF (ABF),
\begin{equation*}
ABF := \frac{1}{2} \left| \frac{Z^n_k( \y )}{Z( \y )} - 1 \right|.
\end{equation*}
To do that, they bound the \textit{expected} ABF (the EABF),
$$
EABF = \int \frac{1}{2} \left| \frac{Z^n_k( \y )}{Z( \y )} - 1 \right| Z( \y ) dy ,
$$
in terms of estimates on the error in the numeric FM. In Theorem \ref{teo:eabf2}, we state the main result of \cite{christen2018}, and the following are the assumptions required.

Assume that we observe a process $ \y = (y_1, \ldots , y_m )$ at locations $z_1, \ldots , z_m \in D$. This is a general setting, to include PDEs and other IPs, in which the domain $D$ may include, for example, space and time: $z_i = (x_i, t_i)$. That is, $z_i$ is an observation at coordinates $x_i$ and at time $t_i$.

\begin{assumption}\label{assp:1}
\textit{Assume that, for all $ \y \in \R^m$, the observation model $f_o( \y | \eta ) $ is uniformly Lipschitz continuous on $\eta $, and for $ \y \in \R^m$, $f_o( y | \eta )$ is bounded. Moreover, the FMs $\cH \circ \cF$ and $\cH \circ \cF^{\alpha(n)}$ are continuous.}
\end{assumption}

\begin{assumption}\label{assp:2}
\textit{Assume a global error control of the numeric FM as
\begin{equation}\label{eqn:global_error1}
||  \cH(\cF(\theta)) - \cH( \cF^{\alpha(n)}(\theta) ) ||_\infty < K^{\alpha(n)}.
\end{equation}
Note that this is a global bound, valid for all $\theta \in \Theta$, and includes already the observational operator. That is, it is a global bound, but is only a statement at the locations $\cH_i$'s where each $y_i$ is observed.}
\end{assumption}

\begin{theorem}\label{teo:eabf2}
(Capistrán et al. \cite{christen2018}) With assumptions~\ref{assp:1}--\ref{assp:2}, and assuming independent data, $\y$, arising from a location-scale family, with scale parameter $\sigma^2$ and location parameter $\eta = \cH(\cF(\theta)) = (\cH_1(\cF(\theta)), \ldots , \cH_m(\cF(\theta))) ^T$, namely
\begin{equation} \label{eq:ModelLocScal}
f_o( \y | \eta ) = \prod_{i=1}^m \sigma^{-1} \rho\left( \frac{y_i - \eta_i}{\sigma} \right),
\end{equation}
with $\rho$ a bounded $C^1$ symmetric Lebesgue density in $\R$, with $\int_{-\infty}^{\infty} x^2 \rho(x) dx = 1$, then
\begin{equation}\label{eqn:EABF2}
EABF < \rho(0) \frac{K^{\alpha(n)}}{\sigma}  m .
\end{equation}
\end{theorem}

Note that model \eqref{eq:ModelLocScal} can be written as
\begin{equation}
y_i = \cH_i(\cF(\theta)) + \sigma \varepsilon_i, \quad i = 1, \ldots, m,
\end{equation}
where each $\varepsilon_i$ has zero mean and unit variance, and its probability distribution function belongs to the location-scale family.

\subsection{Choosing a solver discretization} %a step size in the numerical FM

The bound obtained in Theorem \ref{teo:eabf2} allows deciding what precision to run the solver. The idea is to keep the EABF below a small threshold (e.g., $\frac{1}{20}$) so that the BF is close to 1, and the difference between the numeric and the theoretical model is ``not worth more than a bare mention''  \cite{KASS1995, Jeffreys61}. If we let the $EABF < b$, we need the numerical error in the FM in (\ref{eqn:global_error1}) satisfies %in (\ref{eq:global_error1})
\begin{equation} \label{eq:Bound}
K^{\alpha(n)} < \frac{\sigma}{m} \frac{b}{\rho(0)}.
\end{equation}

Note that, in practice, there is no need to establish the global bound \eqref{eqn:global_error1} theoretically, but rather by a careful strategy for actual global error estimation. In most cases, the posterior distribution is sampled using MCMC, which requires the approximated likelihood at each of many iterations; an automatic process of global error estimation and control will be necessary to comply with \eqref{eqn:global_error1}. We propose a MCMC algorithm with refinement to assure to comply the global bound \eqref{eqn:global_error1} for all $\theta$ in the parametric space of interst.

Assume we have an algorithm to simulate from the posterior distribution. Algorithm \ref{alg:MCMC_ada} describes a strategy for incorporating the bound \eqref{eq:Bound} and the after-the-fact error estimate, proposed in Section~\ref{sec:ErrorAnaylsis}, to control the error in the posterior distribution.

The basic idea of Algorithm \ref{alg:MCMC_ada} is to start with a relatively large step size (e.g., $h = 0.1$), and the step size in time is established to keep the stability condition $k = \frac{3}{4}h^{2}$. At each iteration, $\theta_i$, of the MCMC, the FM, $\cF^{h}(\theta_i)$, is computed, including the after-the-fact error estimate $\hat{K}^{h}_{\theta^{i}}$, using Algorithm \ref{alg:num_est}. If the error in the FM does not comply with the bound in \eqref{eq:Bound}, then run the solver again reducing the spatial step size by half. In the process, we assure \eqref{eqn:global_error1} for all $\theta \in \Theta$.

\begin{algorithm}[ht] 
	\caption{Numerical refinement for the FM in the MCMC algorithm}
	\label{alg:MCMC_ada}

\textit{Step 1: }{\bf Initialization:}\\
\begin{itemize}
\item  Spatial step size $h$ (large)
\item Standard error $\left( \sigma \right)$, sample size $\left( m \right)$, and $\rho ( 0 )$
\item Calculate the error bound $B = \frac{\sigma}{m} \frac{b}{\rho(0)}$, with a tolerance $b$, we suggest $b = \frac{1}{20}$

\item Initial value for the parameter, $\theta^{0}$
\item MCMC length M (number of simulations)
\end{itemize}

\textit{Step 2}. For $i = 1, 2, \ldots, M$:

\hspace{10mm} \textit{Step 3.} Compute the FM, $\cF^{h}(\theta^{i-1})$, and the error estimation $\hat{K}^{h}_{\theta^{i-1}}$, using Algorithm \ref{alg:num_est}

\hspace{10mm} \textit{Step 4.} If $\hat{K}^{h}_{\theta^{i-1}} > B$

	\hspace{30mm} - Set $h = h/2$
	
	\hspace{30mm} - Return to \textit{Step 3}
	
\hspace{22mm} Else

	\hspace{30mm} - Simulate $\theta^{i}$ with some MCMC algorithm

\textit{Step 5:} Output: $(\theta^{0}, \theta^{1}, \ldots, \theta^{M})$

\end{algorithm}
\vspace{1cm}

%############### Numerical examples ###########################
\section{Numerical examples}\label{sec:NumericExamples}
In this section, we use the three previous examples to show the performance of Algorithm \ref{alg:MCMC_ada}, in the solution of the corresponding BIP, using simulated data sets.

We simulate data as follows. The (synthetic) observations, $\y = (y_1,\ldots, y_m)$, are generated under an independent Gaussian model
$$
f_o( \y | \mathbf{\eta}) = \prod_{i=1}^m \sigma^{-1} \rho\left( \frac{y_i - \eta_i}{\sigma} \right)
$$
with $\rho(x) = \frac{1}{\sqrt{2\pi}} e^{-\frac{x^2}{2}}$, i.e.,
\begin{equation}
y_i = \cH_i(\cF(\theta)) + \sigma \varepsilon_i, \label{eq:modelerror}
\end{equation}
where the $\varepsilon_i$'s are independent and identically distributed as $\cN(0, 1)$, $\theta \in \Theta \subset \mathbb{R}^d$ is a vector of unknown parameters, and $\cF( \theta )$ represents the FM. In all our examples, we consider the variance, $\sigma^2$, to be known.

The solution of (\ref{eq:PDE}) with its initial and boundary conditions defines our FM, and we take $\cH (x) = x$ as the observation operator. We consider the BIP to estimate the parameter $\theta$ given observations $\cH_i( \cF(\theta) ) = u(x_{i}, t_1, \theta )$ at some points in space $x_i$, for $i=1,2, \ldots, m$ and at a fixed time $0 < t_1 \leq \tau$. We let the system evolve until time $t_1$ and then observe it at the spacial locations $x_i$'s. The resulting observations are 

$y_i = u( x_i, t_1, \theta) + \sigma \varepsilon_i; \varepsilon_i \sim \cN(0,1)$.

The IP will be treated as a statistical inference problem under a Bayesian approach, setting a prior distribution on the unknown parameter.

The IP will be treated as a statistical inference problem under a Bayesian approach, setting a prior distribution, $\pi_\Theta(\theta)$, on the unknown parameter, to obtain the posterior distribution, $\pi_{\Theta | \Y}( \theta | \y)$, from which all the required inferences are drawn \cite{FOXetAl2013, stuart2010inverse}. The implementation was done using MCMC, through a generic MCMC algorithm, called the t-walk \cite{christen2010general}.

Note that considering independent data with a Gaussian model, the first part of assumption \ref{assp:1} is right, and we only require to verify that $\cH \circ \cF$ and $\cH \circ \cF^{\alpha(n)}$ are continuous. Indeed, the latter is true if the observation operator is the identity. With this scheme, all the necessary assumptions for Theorem \ref{teo:eabf2} are satisfied. And, in this case, $\rho( 0 ) = \frac{1}{\sqrt{ 2 \pi }}$ and the threshold $B := \frac{\sigma}{m} \frac{b}{\rho(0)}$ in \eqref{eq:Bound}, for the numerical error in the FM, is

\begin{equation}
B =  \frac{\sigma}{m} \frac{\sqrt{ 2 \pi }}{20}.
\end{equation}

\begin{example}[Inverse Problem - Fisher's equation]\label{exam:Fisherb}
We consider the BIP to estimate $\theta = r$ in Fisher's equation of Example \ref{ex:fisher}, given measurements of $\cH_i( \cF(\theta) )$ at time $t_{1} = 0.4$. The synthetic data are simulated with the error model (\ref{eq:modelerror}), using the analytical solution for the FM, and the following parameters: $\theta = 4$ and $\sigma = 0.007$, to maintain a 0.01 signal-to-noise ratio. The solution of (\ref{eq:Exa1}) with its initial and boundary conditions defines our FM. We consider $n = 8$ observations at locations $x_{i}$ regularly spaced between $0$ and $1$. The data are plotted in Fig.~\ref{fig:data-hist-Fisher} (a).

Considering a tolerance $b = \frac{1}{20}$ in \eqref{eq:Bound} and with the standard error and sample size used, the error bound for the FM is $B  = 1.1 \times 10^{-4}$. We require a prior distribution, $\pi\left(\cdot\right)$, for the parameter $\theta$; it is assumed $\theta \sim \text{Gamma}(\alpha_1,\beta_1)$ with all known hyperparameters. Regarding the numerical solver, we begin with a (relatively) large step size, $h = 0.05$, and the step size in time is established to keep the stability condition $k=\frac{3}{4} h^{2}$. Then, we start the Algorithm \ref{alg:MCMC_ada}. For $h = 0.01$, the bound is achieved for all iterations.

We compare the posterior distributions using the numerical FM vs. the exact FM, with 200,000 iterations of the t-walk; the histogram is reported with 150,000 samples since the first (burn-in) 50,000 are discarded. The results are shown in Fig.~\ref{fig:data-hist-Fisher} (b) and Table \ref{tab:Comp_T_Num}. The differences observed in both results may be attributed to the Monte Carlo sampling.

\begin{figure}[ht]
\centering
\includegraphics[scale=0.3]{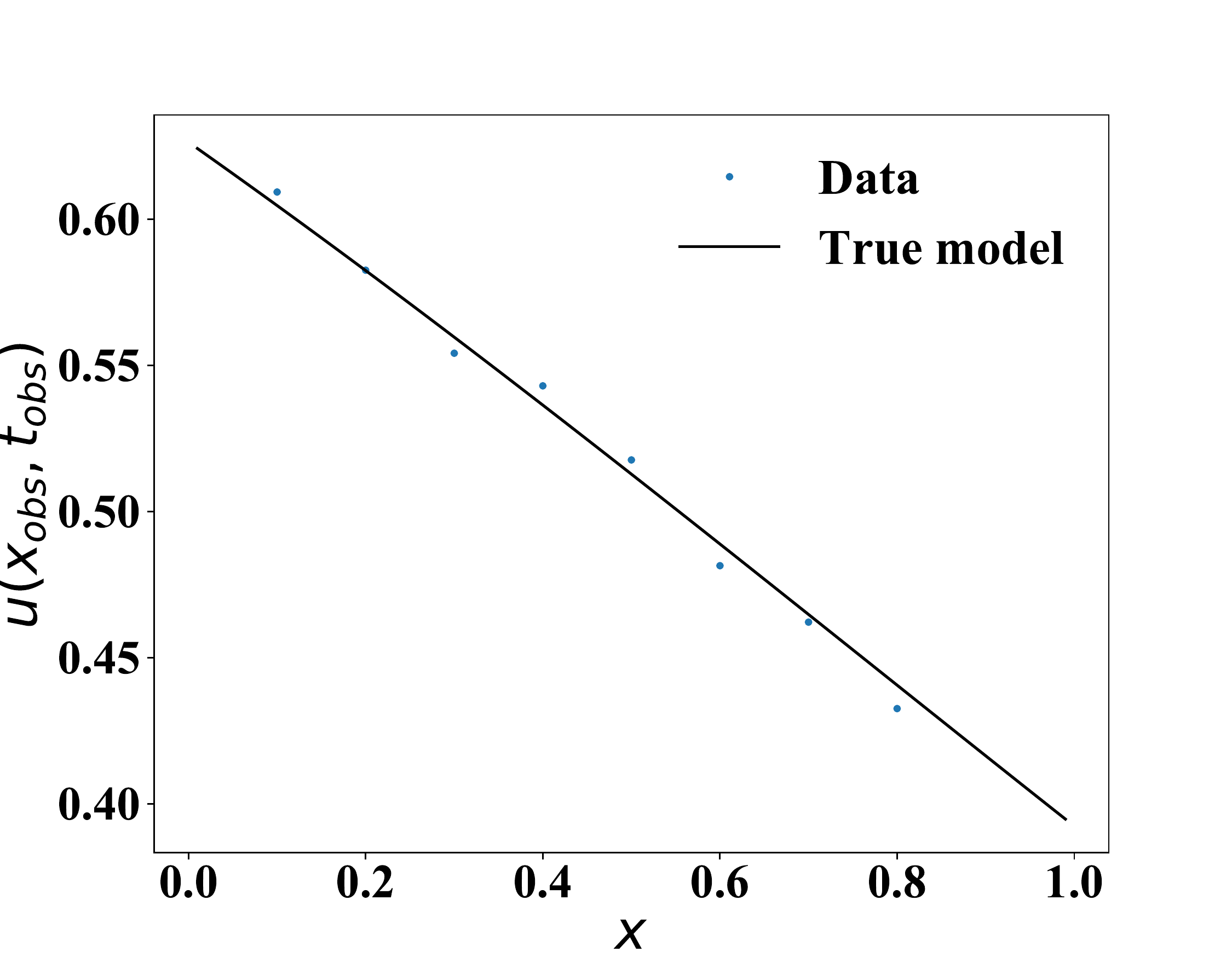}\includegraphics[scale=0.3]{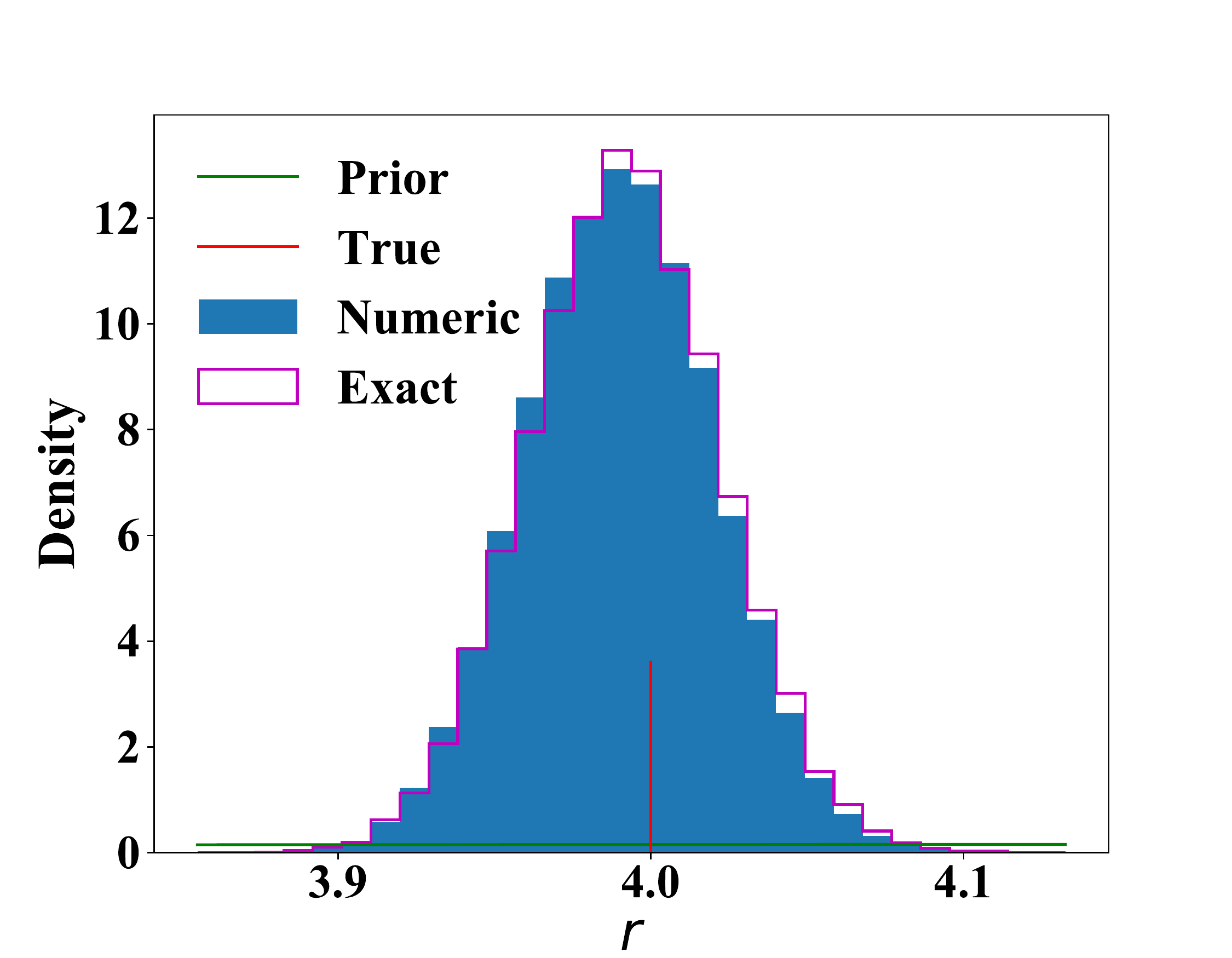}\\
(a)\hspace*{200pt}(b)
\caption{(a) Fisher equation example data (blue points) and true model (black line), considering $r = 4$. (b) Comparison between numerical (blue) and theoretical (magenta) posterior for parameter $r$. The green line represents the prior distribution.
\label{fig:data-hist-Fisher}}
\end{figure}
\end{example}

\begin{example}[Fitzhugh-Nagumo equation]
For this example, the IP is to estimate $\theta=a$ in the Fitzhugh-Nagumo equation of Example \ref{ex:Fitz-Nag}, given measurements $\cH_i( \cF(\theta) )$ at time $t_1 = 0.3$. The synthetic data are simulated with the error model (\ref{eq:modelerror}), using the analytical solution for the FM, and the following parameters: $\theta = 0.3$ and $\sigma = 0.007$. The solution of (\ref{eq:Exa2}) with its initial and boundary conditions defines our FM. We consider $n=8$ observations at locations $x_{i}$ regularly spaced between 0 and 1. The data are plotted in Fig.~\ref{fig:data-hist-Fit-Nag} (a).

To be able to get the posterior distributions, we assume that $\theta\sim\text{Gamma}(\alpha_2,\beta_2)$ with all known hyperparameters. With the standard error and sample size used, and considering a tolerance $b = \frac{1}{20}$ in \eqref{eq:Bound}, we have that the error bound for the FM is $B = 1.1 \times 10^{-4}$. Regarding the numerical solver, we begin with a step size, $h = 0.1$, and the step size in time $k=\frac{3}{4}h^{2}$. Then, we start the Algorithm \ref{alg:MCMC_ada}. For $h = 0.0125$, the bound is achieved for all iterations.

We compare the posterior distributions using the numerical FM vs. the exact FM, with 200,000 iterations of the t-walk; the histogram is reported with 150,000 samples since the first (burn-in) 50,000 are discarded. The results are shown in Fig.~\ref{fig:data-hist-Fit-Nag} (b) and in Table \ref{tab:Comp_T_Num}. The differences observed in both results may be attributed to the Monte Carlo sampling.

\begin{figure}[ht]
\centering
\includegraphics[scale=0.3]{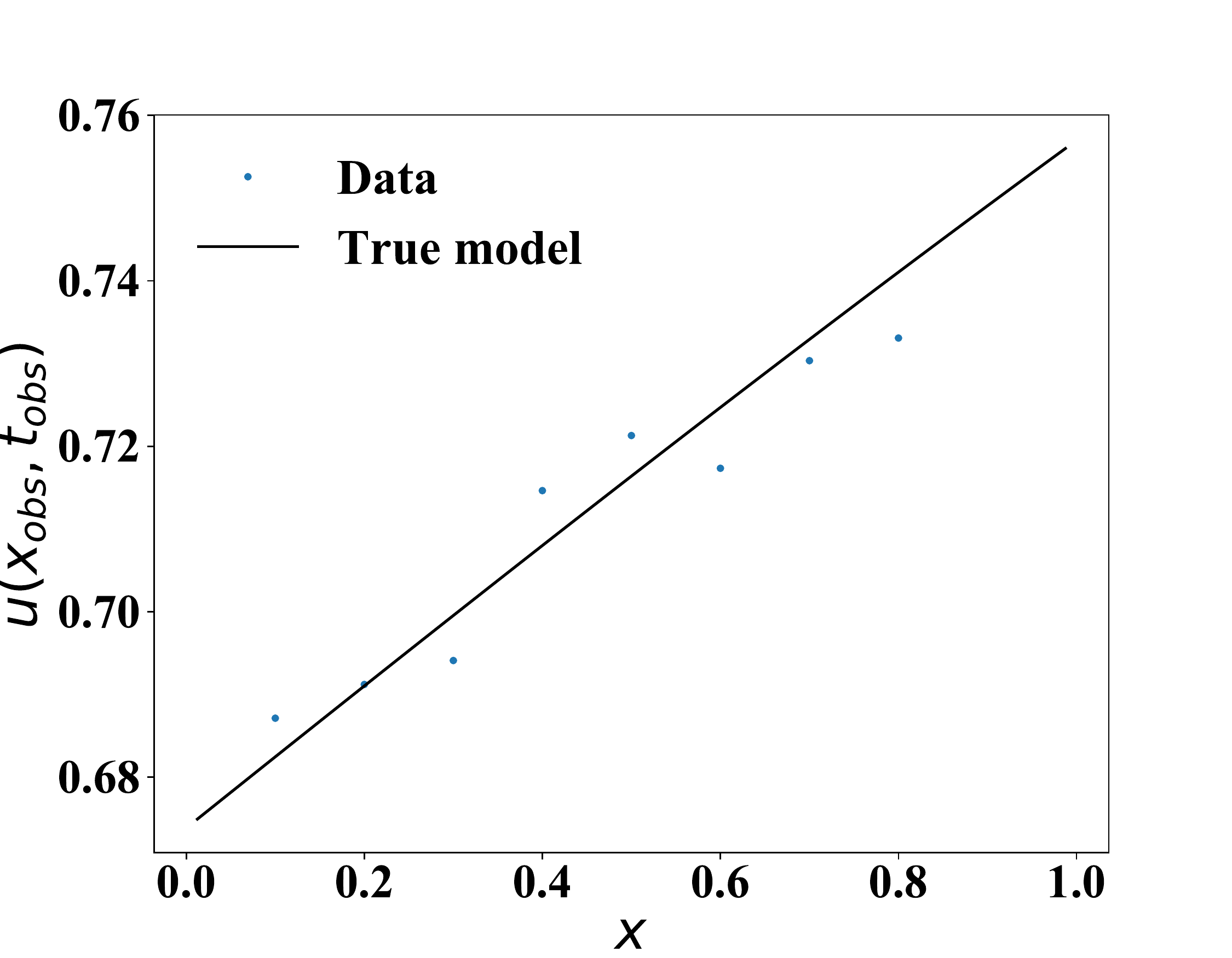}\includegraphics[scale=0.3]{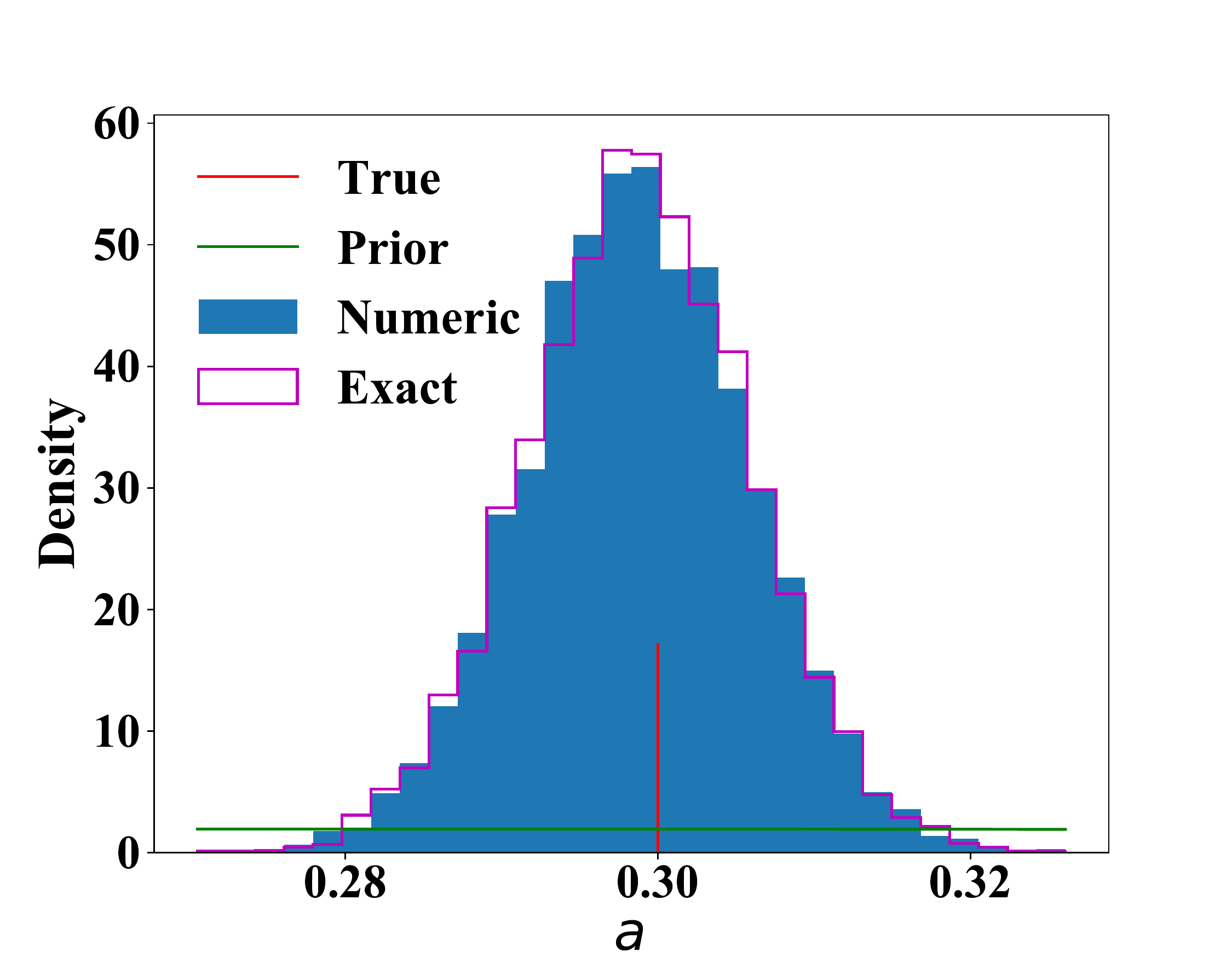}\\
(a)\hspace*{200pt}(b)\\
\caption{(a) Fitzhugh-Nagumo equation example data (blue points) and true model (black line), considering $a=0.3$. (b) Comparison between numerical (blue) and theoretical (magenta) posterior for parameter $a$. The green line represents the prior distribution.
\label{fig:data-hist-Fit-Nag}}
\end{figure}

\end{example}

\begin{example}[Burgers-Fisher equation]

For this example, the IP is to estimate $\theta=(r,s)$ in the Burgers-Fisher equation of Example \ref{exam:Fisherb}, given measurements $\cH_i( \cF(\theta) )$ at time $t_1 = 0.2$. The synthetic data are simulated with the error model (\ref{eq:modelerror}), using the analytical solution for the FM, and the following parameters: $\theta=(4.5,5.5)$ and $\sigma=0.05$. The solution of (\ref{eq:Exa3}) with its initial and boundary conditions defines our FM. We consider $n = 10$ observations at locations $x_i$ regularly spaced between 0 and 1. The data are plotted in Fig.\ref{fig:data-hist-Burger} (a).

To get the posterior distributions, we assume independent priors between the parameters of the model. We assume $r \sim \text{Gamma} \left( \alpha_r, \beta_r \right)$ and $s \sim \text{Gamma} \left( \alpha_s, \beta_s \right)$ with all known hyperparameters. With the standard error and sample size used, and considering a tolerance $b = \frac{1}{20}$ in \eqref{eq:Bound}, we have that the error bound for the FM is $B = 6\times10^{-4}$. Regarding the numerical solver, we begin with a step size $h = 0.1$, and the step size in time $k=\frac{3}{4}h^2$. Then, we start the Algorithm \ref{alg:MCMC_ada}. For $h = 0.0017$, the bound is achieved for all iterations.

We compare the posterior distributions using the numerical FM vs. the exact FM, with 200,000 iterations of the t-walk; the histogram is reported with 150,000 samples since the first (burn-in) 50,000 are discarded. The results are shown in Fig.~\ref{fig:data-hist-Burger} (b)--(c) and in Table \ref{tab:Comp_T_Num}. The differences observed in both results may be attributed to the Monte Carlo sampling.

\begin{figure}[ht]
\centering
\includegraphics[scale=0.3]{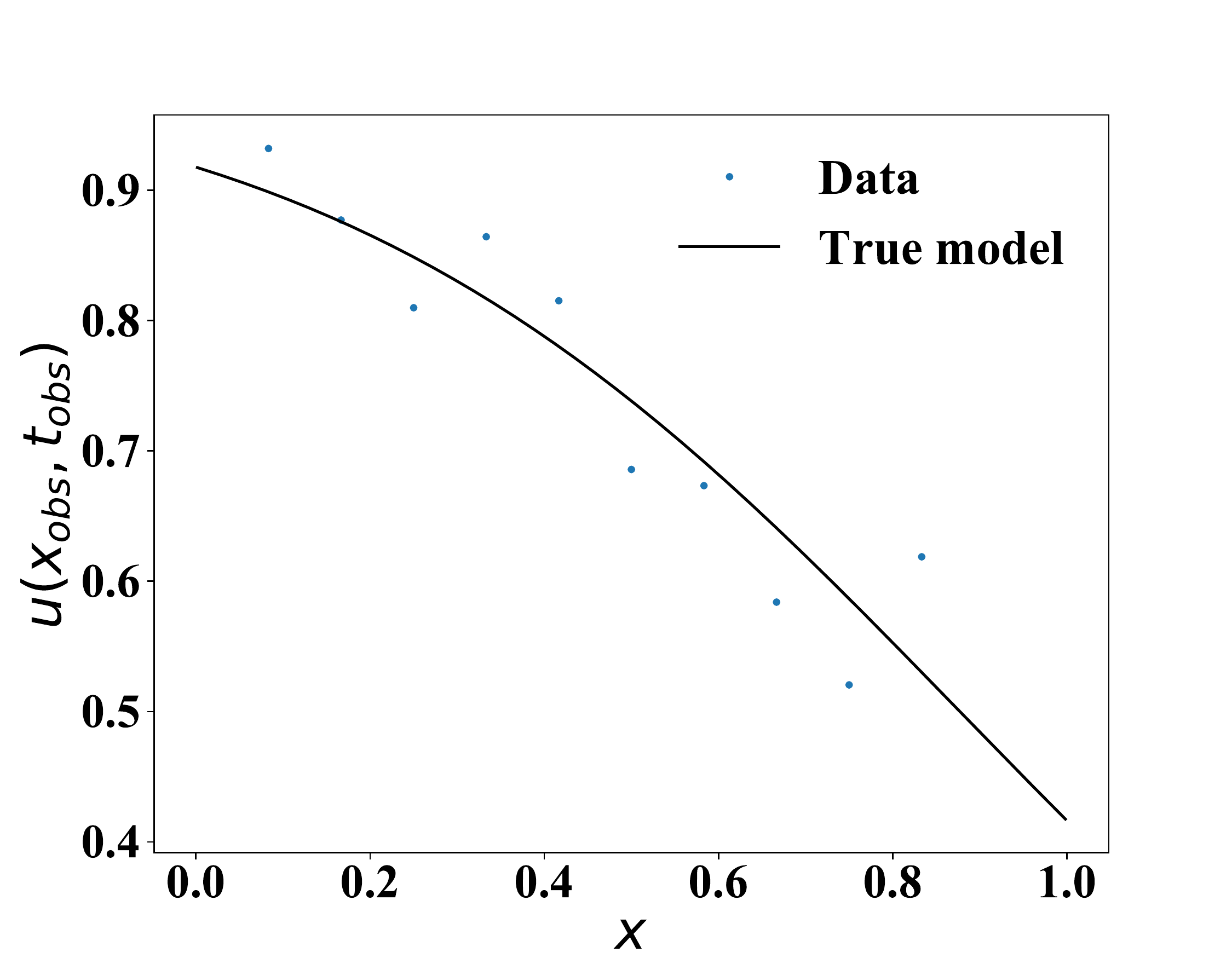}
\includegraphics[scale=0.3]{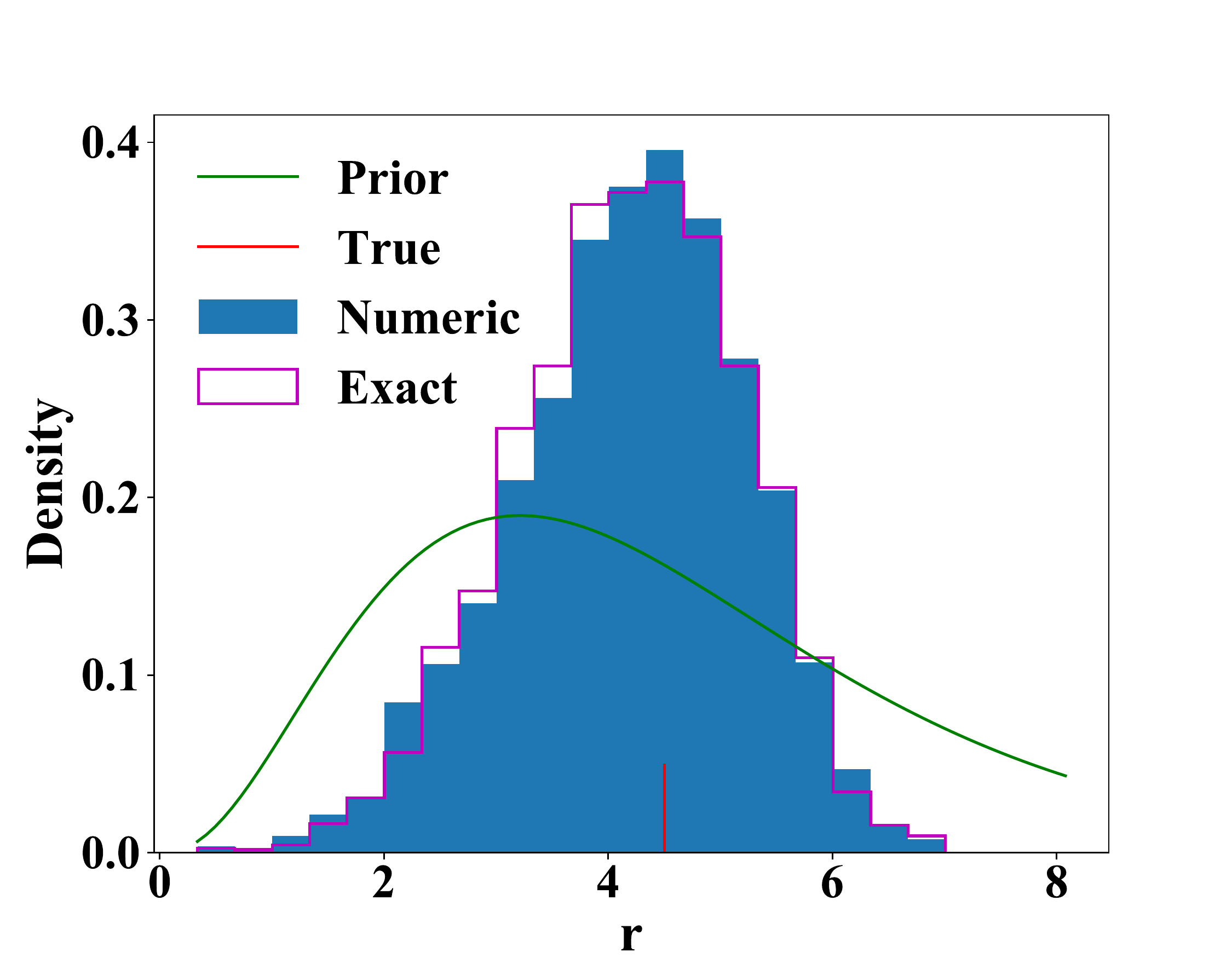}\\
(a)\hspace*{200pt}(b)\\
\includegraphics[scale=0.3]{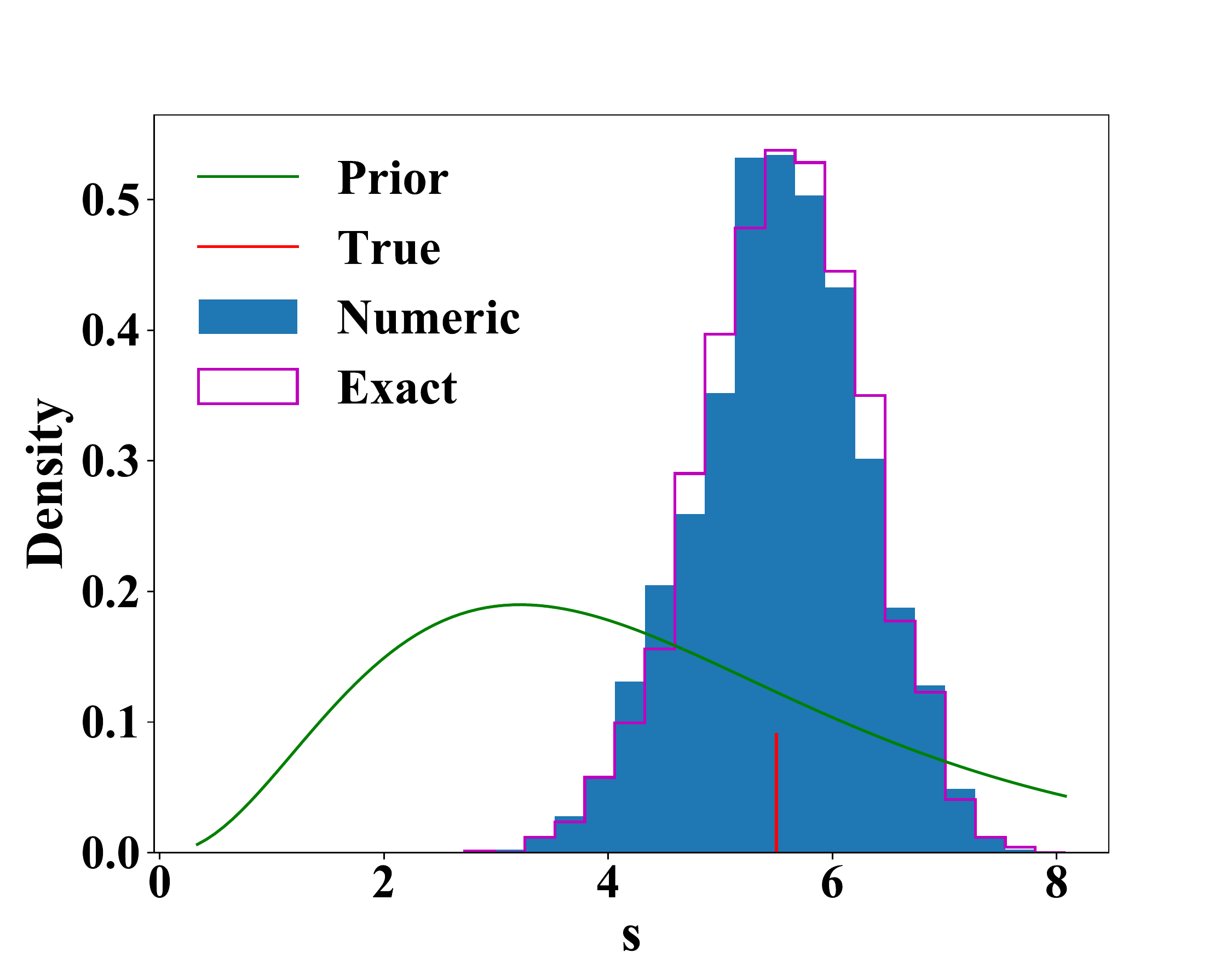}\\
(c)
\caption{(a) Burgers-Fisher equation example data (blue points) and true model (black line), considering $\theta=\left(4.5,5.5\right)$. Histogram from the numerical (blue) and theoretical (magenta) posterior distribution for: (b) parameter $r$ and (c) parameter $s$. The green line represents the prior distribution.
\label{fig:data-hist-Burger}}

\end{figure}

\begin{table}
\centering
\caption{Comparison of the posterior mean (PM) of each parameter using the exact and the numeric FM.\label{tab:Comp_T_Num}}
\begin{tabular}{c c | c | c c}

& \textbf{Example 1} & \textbf{Example 2} & \textbf{Example 3} & \\
\hline \hline
Parameter    & r      & a      & r      & s      \\
\hline
True         & 4      & 0.3    & 4.5    & 5.5    \\
PM-Exact     & 3.9915 & 0.2988 & 4.1813 & 5.5476 \\
PM-Numeric   & 3.9916 & 0.2989 & 4.1859 & 5.5274 \\
\hline
\end{tabular}
\end{table}
\end{example}

As seen in  Figs.~~\ref{fig:data-hist-Fisher} (b), \ref{fig:data-hist-Fit-Nag} (b), and \ref{fig:data-hist-Burger} (b)--(c), and Table \ref{tab:Comp_T_Num}, the histograms and the posterior means obtained with the numerical and the exact FM are practically identical. The small differences observed in both results may be attributed to the effect of generating approximate samples from the posterior distribution using MCMC methods.

\section{Conclusion} \label{sec:Con}
This paper proposed an error estimation for a class of partial differential equations motived by its application in the uncertainty quantification area. Our error estimation allows us to apply the results obtained in \cite{christen2018} for controlling the error in the respective numerical posterior for inverse problems that the forward mapping involves a semi-linear evolution PDE. 

We presented three workout examples; in all cases, the numerical error in the posterior was successfully controlled, which led to a negligible increase in accuracy if the exact FM is considered. This, in turn, may result in CPU time save, as cheaper/rougher solvers are used. 

Although two numerical solutions are required for the error estimation, the added computational effort can be reduced to result equivalent to solving the PDE conventionally (on a single mesh) since evaluating the solution in two different meshes may be easily parallelized. 

For future work, we plan to extend the method used for computing the error estimation to nonlinear evolution differential equations, but some consideration about the stability of the solution and the convergence orders needs to be added.

%\section*{Acknowledgements }
%
%MLDT, JCML, MAC and JAC are partially funded by CONACyT CB-2016-01-284451, RDECOMM and ONRG grants. JCML thanks CONACyT for his PhD scholarship at CIMAT. HH was funded by The Academy of Finland Centre of Excellence of Inverse Modelling and Imaging, decision number 312122.

\newpage
\bibliographystyle{unsrt}
\bibliography{References}

\newpage
\appendix
\section{Details of the numerical solution} \label{sec:ap_ns}
For the reader's convenience, here we describe in detail the numerical method introduced in Section \ref{sec:SolNum}. 

To solve the PDE in Eq.~(\ref{eq:PDE}),  we start by  separating the function $F$ into a linear ($\sL$) and a nonlinear ($\sN$) component and rewriting  the Eq.~(\ref{eq:PDE}) in the form
\begin{equation}\label{eq:PDE_LN1}
\dot{u} = u''+ \sL[u,u'] + \sN[u,u'].
\end{equation} 

The nonlinear operator $\sN$ is approximated with a Taylor series, assuming that the difference $u_{i+1,\cdot} - u_{i,\cdot}$ and all its spatial derivatives are small, hence

\begin{equation}\label{eq:Taylor1}
\sN[u_{i+1,\cdot}, u_{i+1,\cdot}'] \approx \sN[u_{i,\cdot}, u'_{i,\cdot}] + \phi_{0,i} [u_{i,\cdot}, u'_{i,\cdot}]\cdot(u_{i+1,\cdot} - u_{i,\cdot}) + \phi_{1,i}[u_{i,\cdot}, u'_{i,\cdot}]\cdot(u'_{i + 1,\cdot} - u'_{i,\cdot}),
\end{equation}
where $u_{i,\cdot} := u(x_{i},t)$ is the solution of  Eq.~(\ref{eq:PDE}) evaluated in $(x_{i},t)$  and  $\phi_{k,i}[u_{i,\cdot}, u_{i,\cdot}' ] := \frac{\partial^k \sN}{\partial u^{(k)}}[u_{i,\cdot}, u^{(k)}_{i,\cdot}],$  $k=0,1.$

Substituting Eq.~(\ref{eq:Taylor1}) into Eq.~(\ref{eq:PDE_LN1}), we get

\begin{equation}\label{eq:pde1}
\dot{u}_{i+1,\cdot} = u''_{i + 1, \cdot} + \sL[u_{i + 1,\cdot}, u'_{i+1,\cdot}] + \sN[u_{i,\cdot}, u'_{i,\cdot}] +\phi_{0,i}[u_{i,\cdot}, u_{i,\cdot}' ] \cdot(u_{i+1,\cdot} - u_{i,\cdot}) +\phi_{1,i}[u_{i,\cdot}, u_{i,\cdot}' ] \cdot(u'_{i + 1,\cdot} - u'_{i,\cdot}), 
\end{equation}
for $i = 1,...,N-2$.

 Now, the spatial partial derivatives are approximated using the central difference formula (\ref{eq:df_1})-(\ref{eq:df_0}). We write in matrix form the central differences approximations

\begin{equation}
\V_h' = \frac{1}{2h} \A_x \V_h + \C_x,\label{eq:first_der1}
\end{equation}
and
\begin{equation}
\V_h'' = \frac{1}{h^2}  \A_{xx} \V_h + \C_{xx},\label{eq:second_der1}
\end{equation}
where $V_h = (v_{1,\cdot}, v_{2,\cdot}, \ldots,v_{N-1,\cdot})^T$ approximates $\U = (u_{1,\cdot}, u_{2,\cdot}, \ldots,u_{N-1,\cdot})^T,$

\begin{alignat*}{2}
\mathbf{A}_{x} & = 
\left(
\begin{array}{ccccc}
0 & 1 & 0 & \ldots & 0\\
-1 & 0 & 1 & \ldots & 0\\
0 & -1 & 0 & \ldots & \vdots\\
\vdots & \vdots & \vdots & \ddots & 1\\
0 & \ldots & 0 & -1 & 0
\end{array}
\right)_{(N-2)\times(N-2)}, \quad & \C_x  & =\frac{1}{2h}\left(
\begin{array}{c}
-u_{0, \cdot}\\
0\\
\vdots\\
0\\
u_{N, \cdot}
\end{array}
\right)_{(N-2)\times1},
\end{alignat*}

\begin{alignat*}{1}
\A_{xx} & = \left(
\begin{array}{ccccc}
-2 & 1 & 0 & \ldots & 0\\
1 & -2 & 1 & \ldots & \vdots\\
0 & 1 & -2 & \cdots & \vdots\\
\vdots & \vdots & \ddots & \ddots & 1\\
0 & \ldots & 0 & 1 & -2
\end{array}\right)_{(N-2)\times(N-2)}, \text{ and} \quad
\C_{xx} = \frac{1}{h^2}
\left(
\begin{array}{c}
u_{0, \cdot}\\0\\\vdots\\0\\u_{N, \cdot}
\end{array}
\right)_{(N-2)\times1}.
\end{alignat*}

Now, $u_{0,\cdot}'$ is approximated with the forward difference scheme (\ref{eq:df_0}), which leaves us

\begin{equation}
\Vh_h' = \frac{1}{2h} \Ah_x \V_h + \Ch_x,\label{eq:first_der01}
\end{equation}

\begin{alignat*}{1}
\Ah_{x} & = \left(
\begin{array}{ccccc}
2 &0 & 0 & \ldots & 0\\
0 & 1 & 0 & \ldots & \vdots\\
-1 & 0 & 1& \cdots & \vdots\\
\vdots & \vdots & \ddots & \ddots & 1\\
0 & \ldots & 0 & 1 & 0
\end{array}\right)_{(N-2)\times(N-2)}, \text{ and} \quad
\Ch_{x} = \frac{1}{2h}
\left(
\begin{array}{c}
-2u_{0, \cdot}\\-u_{0, \cdot}\\\vdots\\0\\u_{N, \cdot}
\end{array}
\right)_{(N-2)\times1}.
\end{alignat*}
where $\Vh_h = (v_{0, \cdot},  v_{1,\cdot}, \ldots,v_{N-2,\cdot})^T.$

Finally, substituting the approximate derivatives in Eqs.~(\ref{eq:first_der1})--(\ref{eq:first_der01}) into Eq.~(\ref{eq:pde}), joint with the boundary conditions given in Eq.~(\ref{eq:bound_condi}), we get the following semi-discrete differential equation: 

\begin{eqnarray} \label{eq:sys1}
\dot{\V}_h  &=& \frac{1}{h^2} \A_{xx}\V_h + \C_{xx}+ \sL \left( \V_h, \frac{1}{2h} \A_x \V_h + \C_x \right) +\sN \left(\Vh_h, \frac{1}{2h}\Ah_x \V_h + \Ch\right)  \\  \nonumber
& & + \mathbf{\Phi_0}\left[\Vh, \frac{1}{2h} \Ah_x \V_h + \Ch\right] \cdot \left(\V_h - \Vh_h\right)\\  \nonumber
& & + \mathbf{\Phi_1} \left[\Vh_h, \frac{1}{2h} \Ah_x \V + \Ch \right] \cdot \left( \frac{1}{2h}(\A_x -\Ah_x) \V_h + (\C_x - \Ch_x) \right),
\end{eqnarray}
where $\mathbf{\Phi_{k}}[\Vh, \Vh'] = ( \phi_{k,0}[ v_{0, \cdot}, v_{0,\cdot}' ], \phi_{k,1}[v_{1, \cdot}, v_{1,\cdot}' ], \ldots, \phi_{k, N-3}[v_{N-3, \cdot}, v_{N-3, \cdot}' ] )^T.$ 

Note that the right-hand side of equation ( \ref{eq:sys1}) only depends on $\V$ and $t$, due $v_{0, \cdot}$ and $v_{N, \cdot}$ are known (\ref{eq:bound_condi}). Thus, we can write (\ref{eq:sys1}) in a compact form,

\begin{equation} 
\dot{\V}  =  \frac{1}{h^2} \A_{xx}\V + \F(t,\V)
\end{equation}
with

\begin{eqnarray*} 
\F(t,\V) &=& \C_{xx}+ \sL \left( \V_h, \frac{1}{2h} \A_x \V_h + \C_x \right) +\sN \left(\Vh_h, \frac{1}{2h}\Ah_x \V_h + \Ch\right)\\
& &  + \mathbf{\Phi_0}\left[\Vh, \frac{1}{2h} \Ah_x \V_h + \Ch\right] \cdot \left(\V_h - \Vh_h\right)\\
& & + \mathbf{\Phi_1} \left[\Vh_h, \frac{1}{2h} \Ah_x \V + \Ch \right] \cdot \left( \frac{1}{2h}(\A_x -\Ah_x) \V_h + (\C_x - \Ch_x) \right).
\end{eqnarray*}

\section{Stability Considerations} \label{sec:ap_sc}

We briefly describe stability considerations for the DFRK method introduced in Section \ref{sec:SolNum}. 

Let 
\begin{align*}
\mathbf{W}_{\cdot,n+1} & = \mathbf{W}_{\cdot, n} + \left(\frac{b_{1} k} {h^{2}} \mathbf{A}_{xx} \mathbf{W}_{\cdot, n} + kb_{1}\mathbf{F}\left(t_{n}, \mathbf{W}_{\cdot, n} \right)\right) + \sum_{i=2}^{6} \frac{b_{i} k}{h^{2}} \mathbf{A}_{xx}\left(\mathbf{W}_{\cdot, n} + k\sum_{j=1}^{i-1} a_{ij} \mathbf{K}_{j}\right) \\
 & \quad + k\sum_{i=2}^{6}b_{i} \mathbf{F} \left(t_{n} + c_{i} k, \mathbf{W}_{\cdot,n} + k \sum_{j=1}^{i-1} a_{ij} \mathbf{K}_{j}\right)
\end{align*}
the solution of (\ref{eq:PDE}) using the FD-RKCK method (see Section \ref{sec:SolNum}).  To determine the CFL condition, we consider only the pure diffusion.  Thus, the scheme is stable only if $\rho\left(\mathbf{A}\right)\leq1$ \cite{burden2000numerical}, where $$\mathbf{A}=\frac{b_{\text{max}} k}{h^{2}}\mathbf{A}_{xx} \quad b_{\text{max}}=\max_{i}b_{i}.$$

The eigenvalues of $\mathbf{A}$ can be shown to be
$$
\mu_{i} = -4 \lambda \left( \sin\left(\frac{i\pi}{2N} \right)\right)^{2},\quad \text{for} \quad i=1,2,\ldots,N-2,
$$
where $\lambda =b_{\text{max}}  \frac{k}{h^{2}}$. So, the condition for stability consequently reduces to determining if
$$
\text{\ensuremath{\rho\left(\mathbf{A}\right)}}=\max_{1\leq i\leq N-2}\left|-4\lambda\left(\sin\left(\frac{i\pi}{2N}\right)\right)^{2}\right|\leq1,
$$
and this simplifies to
$$
0 \leq \lambda \left( \sin \left(\frac{i\pi}{2N} \right)\right)^{2} \leq \frac{1}{4}, \quad \forall i=1,2,\ldots,N-2.
$$

Stability requires that this inequality condition hold as $h \rightarrow 0$,
or, equivalently, as $N \rightarrow \infty$,

$$
\lim_{N \rightarrow \infty}\left[ \sin \left( \frac{\left(N-1 \right) \pi}{2N} \right) \right]^{2} = 1.
$$

Thus, stability occurs if only if $0 \text{\ensuremath{\leq \lambda \leq}}\frac{1}{4}$.
By definition $\lambda =b_{\text{max}} \frac{k}{h^{2}}$, so this inequality requires that $h$ and $k$ be chosen such that
$$
b_{\text{max}} \frac{k}{h^{2}}\leq\frac{1}{4}.
$$
The method converges to the solution with a rate of convergence $O \left( h^{p} + k^{4} \right)$,
provided $b_{\text{max}}\frac{k}{h^{2}}\leq\frac{1}{4}$. For
the numerical implementation, we take $k = \alpha h^{2}$, with $\alpha = \frac{1}{4b_{\text{max}}}$.

\end{document}